\documentclass[onecolumn, 10pt]{WileyMSP-template}

\usepackage{style_sheet}
\usepackage{amsmath}
\usepackage{amsfonts}
\usepackage{amssymb}
\usepackage{upgreek}
\usepackage{scrextend}
\usepackage{hyperref}
\usepackage{cite}
\usepackage{caption}
\usepackage{comment}
\usepackage{threeparttable}
\usepackage{colortbl}
\usepackage{booktabs}
\usepackage[normalem]{ulem}  

\renewcommand{\d}{\mathop{}\!\mathrm{d}}

\renewcommand{\eqref}[1]{Eq.~(\ref{#1})}
\newcommand{\eqsref}[2]{Eqs.~(\ref{#1}--\ref{#2})}
\newcommand{\eqreftwo}[2]{Eqs.~(\ref{#1}, \ref{#2})}
\newcommand{\secref}[1]{Sec.~\ref{#1}}
\newcommand{\figref}[1]{Fig.~\ref{#1}}
\newcommand{\tabref}[1]{Tab.~\ref{#1}}
\newcommand{\appref}[1]{App.~\ref{#1}}

\usepackage{refcount}

\usepackage{titlesec}

\makeatletter
\newcommand{\startappendices}{%
  \appendix
  \renewcommand{\thesection}{\Alph{section}}%
  \setcounter{section}{0}%
  
  \@addtoreset{equation}{section}%
  \renewcommand{\theequation}{\Alph{section}\arabic{equation}}%
  
  \titleformat{\section}[block]
    {\normalfont\Large\bfseries} 
    {}                            
    {0pt}{Appendix~\Alph{section}:~}
}
\makeatother

\begin{document}

\title{Efficient first-principles inverse design of nanolasers}
\maketitle

\author{Beñat Martinez de Aguirre Jokisch*,}
\author{Alexander Cerjan,}
\author{Rasmus Ellebæk Christiansen**,}
\author{Jesper Mørk,}
\author{Ole Sigmund,}
\author{and Steven G. Johnson.}

\begin{affiliations}
\noindent
Beñat Martinez de Aguirre Jokisch, Rasmus E. Christiansen, Ole Sigmund \\

Department of Civil and Mechanical Engineering, Technical University of Denmark, Nils Koppels Allé, Building 404, 2800 Kongens Lyngby, Denmark. \\
*\textcolor{white}{*}  Email: bmdaj@dtu.dk \\
** Email: raelch@dtu.dk

\vspace{0.5em}
\noindent
Jesper Mørk \\
Department of Electrical and Photonics Engineering, Technical University of Denmark, Ørsteds Plads, Building 343, 2800 Kongens Lyngby, Denmark.

\vspace{0.5em}
\noindent
Beñat Martinez de Aguirre Jokisch, Rasmus E. Christiansen, Jesper Mørk, Ole Sigmund \\
NanoPhoton – Center for Nanophotonics, Technical University of Denmark, Ørsteds Plads, Building 345A, 2800 Kongens Lyngby, Denmark.

\vspace{0.5em}
\noindent
Alexander Cerjan \\
Center for Integrated Nanotechnologies, Sandia National Laboratories, Albuquerque, NM 87185, USA.

\vspace{0.5em}
\noindent
Steven G. Johnson \\
Department of Mathematics, Massachusetts Institute of Technology, Cambridge, MA 02139, USA.
\end{affiliations}



\begin{abstract}

We develop and demonstrate a first-principles approach, based on the nonlinear Maxwell--Bloch equations and steady-state \textit{ab-initio} laser theory (SALT), for inverse design of nanostructured lasers, incorporating spatial hole-burning corrections, threshold effects, out-coupling efficiency, and gain diffusion. The resulting figure of merit exploits the high-$Q$ regime of optimized laser cavities to perturbatively simplify the nonlinear model to a single \emph{linear} ``reciprocal'' Maxwell solve. The consequences for laser-cavity design, and in particular the strong dependence on the nature of the gain region, are demonstrated using topology optimization of both 2d and full 3d geometries.

\end{abstract}

    
\section{Introduction}\label{sec:intro}


    \begin{figure}[hb!] 
    \centering
    \includegraphics[scale=1]{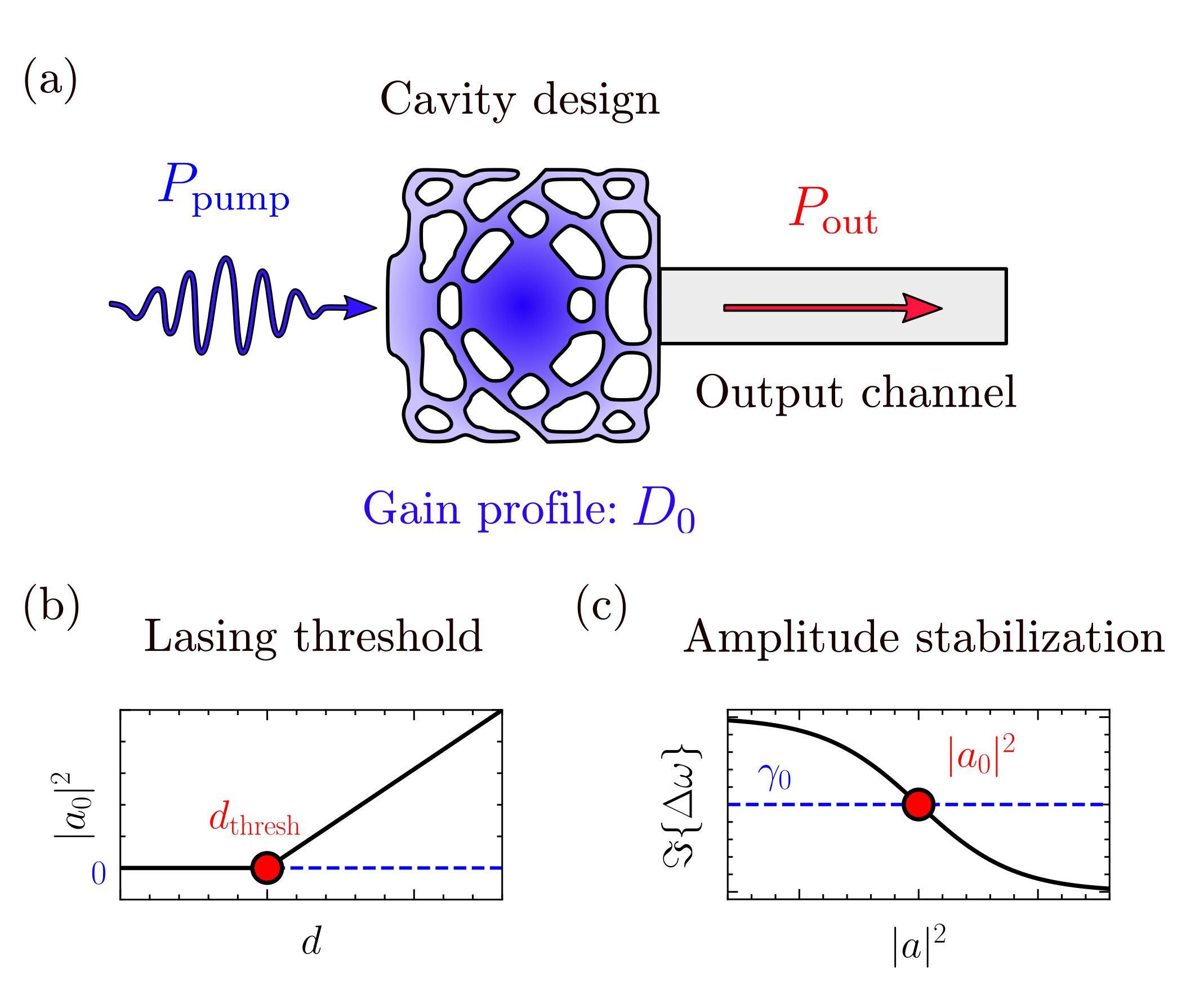} 
\caption{(a) Schematic nanolaser: pump power $P_\text{pump}$ excites a gain profile $D_0$ and a lasing mode that outputs power $P_\text{out}$ into a channel (waveguide), for a cavity (from \secref{sec:benchmark}) that maximizes efficiency $\sim P_\text{out} / P_\text{pump}$. (b) Lasing amplitude $|a_0|^2$ ($\sim\text{energy}$) vs.~pump strength~$d$, lasing above a threshold $d_\text{thresh}$. (c) Change $\Im\{\Delta \omega \}$ in modal gain/loss depends nonlinearly on amplitude $|a|^2$, and stabilizes at $|a_0|^2$, where $\Im\{\Delta \omega\}=$ passive cavity loss $\gamma_0$.}
    \label{fig:schem}
\end{figure}

In this paper, we present a new formulation for efficient inverse design of nanolasers
(\figref{fig:schem}) to maximize efficiency ($\sim \text{output}$ $\text{power} / \text{pump}$ power), taking nonlinear and coupling effects into account, and demonstrate 2d and 3d semiconductor-laser topology optimization (TopOpt, \secref{sec:TopOpt}).
We exploit the fact that optimization evolves the lasing cavity to a high-$Q$ ($\gtrsim
100$, long-lifetime) regime to perturbatively extract hole-burning effects, thresholds, output coupling, and other relevant phenomena from a \emph{single linear} ``reciprocal'' scattering solve excited from the output port for a passive cavity.
Our approach builds on the ``SALT'' (steady-state \textit{ab initio} laser theory) formulation of lasing~\cite{SALT_original, Ge_2010} (\secref{sec:SALT}), which exactly solves the nonlinear Maxwell--Bloch equations for a single-mode lasing steady state, including all nonlinear gain and saturation effects in the Maxwell--Bloch model, but we are able to eliminate expensive nonlinear and linear eigensolves using a new synthesis of SALT, perturbation theory (extending single-pole approximation (SPA)-SALT theory~\cite{Ge_2010}, \secref{sec:SPA}), and temporal coupled-mode theory (TCMT~\cite{phot_crys}, \secref{sec:TCMT}).
Our resulting figure of merit (FOM) is a simple ratio $\sim (\int_{\text{gain}} |\bvec{E}|^2\d\Omega)^3 / \int_{\text{gain}} |\bvec{E}|^4\d\Omega$ (\eqref{eq:eff_nl} in \secref{sec:FOM}) computed from only the linear-model electric field $\bvec{E}$.  We show that this FOM reduces to a more conventional cavity LDOS-like optimization---power emitted into an output channel by a dipole---in the limit of a \emph{point-like} gain region (\secref{sec:single}), but leads to very different (more delocalized) cavity designs and performance for a \emph{distributed} gain region.
We demonstrate TopOpt of this FOM with both 2d and 3d semiconductor-on-insulator designs, including manufacturing constraints, and investigate the effect of gain-region diameter (\secref{sec:benchmark}).  
We also compare to a heuristic FOM $\sim \int_{\text{gain}} |\bvec{E}|^2 \d\Omega$ (\secref{sec:single}), and show that this leads to suboptimal performance (\secref{sec:benchmark}).
Furthermore, we show how to incorporate gain-diffusion effects~\cite{csalt} into TopOpt (e.g., for semiconductor lasers in a free-carrier approximation) at the cost of only two additional damped-diffusion solves (\secref{sec:diffusion_theory}), and show that diffusion favors a disconnected topology of the gain region (\secref{sec:diffusion_results}).
We believe that our results not only demonstrate the importance (and surprising ease) of incorporating sophisticated nonlinear lasing models into the inverse-design process, but also provide a foundation for future extensions, e.g.~to include an explicit model of the pumping process, as discussed in \secref{sec:conclusion}. 
    

The inverse design of nanolasers has conventionally been based on improving generic FOMs that are \emph{related} to lasing, without directly modeling the laser physics or accounting for the spatial distribution of the emitters \cite{xiong2025nanolaser}, such as maximizing the cavity quality factor~$Q$ ($\sim \text{lifetime}/\text{period}$~\cite{phot_crys}). Another such FOM is the local density of states (LDOS), which is proportional to $Q$ divided by a measure of modal volume (the ``Purcell factor'') and is more precisely the total power expended by a dipole source~\cite{OskooiJo13-sources}; this is related to enhancement of light--matter interactions (e.g.~spontaneous emission rate) at a single point~\cite{LDOS_Laser, LDOS_Laser_2, LDOS_Laser_3, bowtie_laser}. For example, high $Q$ factors lead to lower lasing thresholds~\cite{saldutti2023thresholdsemiconductornanolasers} (\secref{sec:SPA}). Nonetheless, and as our results also demonstrate, in nanolaser design a high $Q$ alone is insufficient: while strong temporal confinement (high $Q$) is crucial, the photons emitted by the gain medium must also be efficiently extracted and coupled into an output channel. This concept has been explored in the context of inverse design of light emission, where the optical extraction efficiency~\cite{extraction,Janssen2010, Mayer, Chakravarthi:20, Yao_2022, Chung:22, KimShinLeeParkLeeParkSeoParkParkJang, Shultzman23, schubert2023fouriermodalmethodinverse} ---the (linear-model) output power of some source into some channel(s)---is considered as the optimization FOM. However, for laser design, simple metrics such as the $Q$, LDOS, or extraction efficiency do not account for the precise distribution of the gain medium and its overlap with the optical field  \cite{xiong2025nanolaser} or nonlinear lasing effects (e.g.~gain, saturation).  Below, we show that an LDOS-like FOM only becomes equivalent to a more complete lasing model in the limiting case of a point-like gain region (as shown in \secref{sec:single}). The LDOS is also relevant to other physical phenomena such as sensing, and there have been several research efforts targeting this FOM using inverse-design frameworks like TopOpt~\cite{LDOS_opt_wang, LDOS_opt_liang, LDOS_bounds, albrechtsen_nanometer-scale_2022, LDOS_G}, including experimental realization~\cite{albrechtsen_nanometer-scale_2022}, along with other methods targeting high~$Q$\cite{Ahn2022}.  (TopOpt, reviewed in \secref{sec:TopOpt}, is a gradient-based inverse design framework that finds freeform designs via large-scale optimization where ``every pixel'' is a degree of freedom, and has been successfully applied to many photonic problems~\cite{jensen_topology_2011, Molesky_2018}.)  However, it is known that the result of cavity optimization can be greatly altered if one targets an FOM that is tailored to a phenomenon very different from spontaneous emission at a point, such as Raman scattering~\cite{ChristiansenMi20,PanCh21,YaoVe23}, volume-averaged emission~\cite{Yao_2022}, scintillation effects~\cite{Roques_Carmes_2022}, or harmonic generation~\cite{LinLi16,Mann2023}.

A key enabling factor for first-principles nanolaser design is the recent development of SALT, reviewed in \secref{sec:SALT}~\cite{SALT_original, Ge_2010}, which formulates a tractable model of a full nonlinear lasing problem by transforming the time-domain nonlinear Maxwell–Bloch equations~\cite{haken_Laser_dynamics, PhysRev.134.A1429} into a set of self-consistent, time-independent equations for the lasing steady state. SALT has provided a foundation for many subsequent developments, such as improvements in the theory of laser amplifiers~\cite{I-SALT}, complicated gain media with carrier diffusion~\cite{csalt}, quantum limits on the laser linewidth~\cite{pick}, and random lasers~\cite{random_lasers}.  However, the fact that SALT is based on a nonlinear eigenproblem (arising when a resonance lases) makes it more challenging for inverse design than linear scattering: not only does it have all of the pole-tracking and potential \linebreak non-differentiability of linear eigenvalue/resonance optimization~\cite{LDOS_opt_liang}, but when a pole crosses the real-$\omega$ axis one must also ``turn on'' a nonlinear solver~\cite{EsterhazyLiu14}.  Merely identifying the lasing pole in a differentiable way would be difficult without a good starting guess (perhaps via LDOS optimization), and the nonlinear effects incur additional computational complexity~\cite{EsterhazyLiu14}. Although these issues may be tractable in principle (e.g.~differentiating the nonlinear solve via implicit adjoint methods~\cite{Griewank2008}; such methods have been used to optimize nonlinear steady-state transmission~\cite{Hughes2018}), it would be better to find a more computationally efficient formulation. The reason such an improvement should be possible is the fact that an \emph{optimized} nanolaser is quite special: to be high-efficiency and/or low threshold, the cavity will have a high~$Q$, and it is known that SALT can be simplified by perturbative analysis in this regime~\cite{Ge_2010} (empirically found to be accurate for passive cavity resonances with $Q\gtrsim
100$~\cite{cerjan_2016} and numerically validated in \appref{sec:app_val}), although the previous perturbative ``SPA-SALT'' techniques were not devised with optimization in mind. Here, we show that specifically targeting efficient inverse design allows us to obtain a computationally efficient and simple FOM without sacrificing first-principles physics.

\section{Theoretical framework}

In this section, we introduce the theoretical framework to model nanolasers using SALT in the perturbative single-mode high-$Q$ regime, and combining it with TCMT, we derive a FOM for nanolaser optimization that can be be evaluated with a single reciprocal scattering solve. First, we review laser modeling via SALT (\secref{sec:SALT}), which requires one to solve a system of frequency-domain nonlinear equations to calculate laser properties. Then, we review SPA-SALT (\secref{sec:SPA}) an approximation for the single-mode high-$Q$ cavity limit, where perturbation theory is used to derive analytical expressions for laser properties in terms of the passive cavity (no gain) mode solution. In this limit, the equations simplify to a single eigenvalue problem. Next, we use TCMT (\secref{sec:TCMT}) to derive a FOM (\eqref{eq:eff_nl}) that accounts for the out-coupling into a desired channel, and can be computed with a single reciprocal solve, where instead of considering the emission from the gain medium one excites an output channel and evaluates the linear-Maxwell electromagnetic fields in the cavity. We then compare the FOM to a heuristic generalization of the LDOS for a distributed gain medium, and show that both FOMs are equivalent in the single-emitter limit (\secref{sec:single}). Finally, we define the topology optimization framework (\secref{sec:TopOpt}) to set up an optimization problem that can be solved to maximize the FOM.

\subsection{Review of Steady-State Ab-initio Laser Theory (SALT)}\label{sec:SALT}

In a laser, an external pump excites a gain medium, which then interacts with the optical cavity mode, and emits photons into an output mode through stimulated emission. This is illustrated in \figref{fig:schem}(a) for an example of an inverse designed nanolaser. The gain medium located in the cavity is modeled as an ensemble of two-level atoms interacting with the optical cavity field, which can be described by means of the Maxwell-Bloch equations~\cite{haken_Laser_dynamics, PhysRev.134.A1429, SALT_original}
\begin{equation}
\begin{gathered}
\nabla \times \nabla \times \bvec{E}^{+}-\frac{\varepsilon_c}{c^2} \ddot{\bvec{E}}^{+}=\frac{1}{c^2\varepsilon_0} \ddot{\bvec{P}}^{+}\,, \\
\dot{\bvec{P}}^{+}=\left(-i \omega_a-\gamma_{\perp}\right) \bvec{P}^{+}+\frac{g^2}{i \hbar} \bvec{E}^{+} D \,,\\
\dot{D}=\gamma_{\|}\left(D_0-D\right)-\frac{2}{i \hbar}\left[\bvec{E}^{+} \cdot\left(\bvec{P}^{+}\right)^*-\bvec{P}^{+} \cdot\left(\bvec{E}^{+}\right)^*\right]\,,
\end{gathered}
\end{equation}
where $i$ is the imaginary unit, $\hbar$ is the reduced Planck constant, $g$ is a dipole matrix element, $\bvec{E}^{+}(\bvec{r},t)$ and $\bvec{P}^{+}(\bvec{r},t)$ are the positive frequency components of the electric field and polarizations respectively, $\varepsilon_c(\bvec{r})$ is the complex relative permittivity of the cavity, $\varepsilon_0$ is the vacuum permittivity, $D(\bvec{r},t)$ is the population inversion, $D_0(\bvec{r})$ is the gain profile, which is established through external pumping, $\omega_a$ is the atomic transition frequency, $\gamma_\perp$ is the gain width (polarization dephasing rate), and $\gamma_\parallel$ is the population relaxation rate. Note that these equations rely on the rotating-wave approximation (RWA) \cite{Ge_2010}.

For semiconductor lasers, the population relaxation rate will be much lower than the atomic transition frequency and the dephasing rate ($\omega_a,\gamma_\perp \gg \gamma_\parallel$), which leads to the stationary-inversion approximation (SIA) $\dot{D}\sim 0$~\cite{Ge_2010}. One can then derive the coupled nonlinear SALT differential equations~\cite{Ge_2010}
\begin{equation}\label{eq:SALT}
{\left[ \left( \nabla \times 
 \nabla \times {} \right)-\left(\varepsilon_c(\bvec{r})+\frac{\gamma_{\perp} D(\bvec{r})}{\tilde\omega_i-\omega_a+i \gamma_{\perp}}\right) \left(\frac{ \tilde\omega_i}{c}\right)^2\right] \bvec{E}_i(\bvec{r})=0}\,,
\end{equation}
where  $\omega_i$ and $\bvec{E}_i$ are the complex-valued frequency and mode profile of the $i$-th cavity mode, and the expression for the population inversion is given by
\begin{equation}\label{eq:D_E}
D(\bvec{r})=   \left[1+\sum_{j=1}^N \Gamma_j\,  e_c^{-2}\, \left|\bvec{E}_j(\bvec{r})\right|^2 \right]^{-1} D_0(\bvec{r}) d \, ,
\end{equation}
where $d$ is a scalar-valued pumping strength that we have separated from the spatial pumping profile $D_0(\bvec{r})$,   $\Gamma_j= \gamma_\perp^2 / \left[\gamma_\perp^2+(\omega_j-\omega_a)^2\right]$ is a Lorentzian gain curve evaluated at the frequency $\omega_j$ for a gain medium emitting at the wavenumber $\omega_a$, and $e_c$ is a non-dimensionalization scaling for the fields~\cite{Ge_2010}.

Note that this framework does not explicitly model the physical pumping process: $D_0 d$ is treated here as a fixed input.  As discussed in \secref{sec:conclusion}, an opportunity for future work is to obtain $D_0$ itself by solving an auxiliary system of equations, e.g.~another Maxwell solve for optical pumping.  Also not explicitly included here is the pump power to reach transparency ($d=0$). Although this could be larger than the additional $d > 0$ pump power required to reach/exceed threshold for high $Q$ and low lasing powers, the power to reach transparency is mostly a property of the materials and the pumping process, independent of the lasing mode, so it is not our primary concern in this paper.

\subsection{Single-pole threshold approximation}\label{sec:SPA}

The expression in \eqref{eq:SALT} simplifies when considering a single high-$Q$ lasing mode $L$ ($i=j=L$) that is aligned to the atomic transition $\omega_a=\omega_L$ of the emitters in the gain medium. This approximation, which is accurate for a sufficiently high $Q$ (e.g.~$Q\gtrsim
100$ in previous work~\cite{cerjan_2016} and numerical validation in \appref{sec:app_val}) is also known as the single-pole approximation (SPA-) SALT equation~\cite{Ge_2010}
\begin{equation}\label{eq:SPA_SALT}
{\left[(\nabla \times 
 \nabla \times ) -\left(\varepsilon_c(\bvec{r})-i \Delta \varepsilon_\Im (\bvec{r})\right) \left(\frac{\omega_L}{c}\right)^2\right] \bvec{E}_L(\bvec{r})=0}\,.
\end{equation}
where the change in the dielectric permittivity is given by
\begin{equation}\label{eq:gain_SALT}
    \Delta \varepsilon_\Im (\bvec{r}) =  \frac{D_0(\bvec{r}) d}{1+ e_c^{-2}\left|\bvec{E}_L(\bvec{r})\right|^2}\,.
\end{equation}
\eqref{eq:SPA_SALT}  still involves solving one nonlinear eigenvalue problem. Nevertheless, the expression can be further simplified by perturbatively reducing the equation to linear form. For a high-$Q$ cavity, operating at low powers near threshold, the gain required to balance cavity loss is small. More precisely, in this regime the change in the imaginary part of the dielectric permittivity induced by the active medium will be small ($\propto 1/Q$) with respect to the passive cavity permittivity ($\Delta \varepsilon_\Im \ll \varepsilon_c$), for a high-$Q$ 
passive cavity mode with a complex angular frequency frequency $ \tilde\omega_{1} = \omega_1 - i \gamma_1 $, where the quality factor is defined as  $Q=\omega_1/(2\gamma_1)$. We can then describe the shift in the cavity-mode frequency by using first-order perturbation theory~\cite{Raman_2011}
\begin{equation}\label{eq:pert}
\Delta \tilde\omega_{1} = i \frac{ \omega_1}{2} \frac{\int_{\Omega} \Delta \varepsilon_\Im (\bvec{r})  |\mathbf{e}_{1}(\bvec{r})|^2\,  \d \Omega}{\int_{\Omega} \varepsilon_c(\bvec{r})|\mathbf{e}_{1}(\bvec{r})|^2\,  \d \Omega}\,,
\end{equation}
where $\mathbf{e}_{1}$ is the mode profile of the passive cavity. In the high-$Q$ cavity limit, the lasing mode $\bvec{E}_L$ is approximately proportional to the cavity mode~\cite{Ge_2010} within a $\bigO(1/\sqrt{Q})$ error (proportional to the square root of the radiated power~\cite{phot_crys}). The change in permittivity can then be expressed as
\begin{equation}\label{eq:gain_amp}
        \Delta \varepsilon_\Im (\bvec{r}) \approx \frac{D_0(\bvec{r}) d}{1+ |a_1|^2\,\left( e_c^{-2}|\mathbf{e}_{1}(\bvec{r})|^2  / \int_{\Omega} \varepsilon_c(\bvec{r}) |\mathbf{e}_{1}(\bvec{r})|^2 \d \Omega \right)  }\,,
\end{equation}
where $a_1$ is the amplitude of the lasing mode and the lasing intensity $|a_1|^2$ has units of energy. Note that since $\Delta \varepsilon_\Im \in \mathbb{R}$ this yields a purely imaginary shift in the cavity-mode frequency, which can be used to counteract the losses of the passive cavity, so that the cavity mode starts lasing.

In the low-power regime, for a small amplitude $|a_1|^2$ of the lasing mode, one can expand \eqref{eq:gain_amp} in a Taylor series around $|a_1|^2=0$. As shown in \figref{fig:schem}(b) the amplitude of the lasing mode is zero at threshold, and the system will lase for a pumping strength above threshold $d \geq d_\text{thresh}$. Notice, that we are here considering the approximation where we neglect the spontaneous emission into the lasing mode, which will lead to non-zero power in the cavity mode below the lasing threshold~\cite{saldutti2023thresholdsemiconductornanolasers}. In the zero-amplitude limit, the change in permittivity is linearly related to the pump strength: $ \Delta \varepsilon_\Im (\bvec{r}) = d \, D_0(\bvec{r})$. As shown in \figref{fig:schem}(c), for lasing to occur, the imaginary frequency shift caused by the gain, as expressed in \eqref{eq:pert}, must balance the losses of the passive cavity
\begin{equation}\label{eq:las_cond}
    \Im\{\Delta \tilde \omega_{1}\} - \gamma_1 = 0\,.
\end{equation}
This condition is met at threshold ($d=d_\text{thresh}$), giving an expression for the pumping strength needed to reach threshold
\begin{equation}\label{eq:pump_thresh}
    d_\text{thresh}  = \frac{1}{Q} \frac{\int_{\Omega} \varepsilon_c(\bvec{r})|\mathbf{e}_{1}(\bvec{r})|^2\,  \d \Omega}{\int_{\Omega} D_0(\bvec{r}) |\mathbf{e}_{1}(\bvec{r})|^2\,  \d \Omega}\,.
\end{equation}
As would be expected, the pumping strength needed to reach the lasing threshold is inversely proportional to $Q$ and to the energy confinement in the active medium, which leads to stronger light-matter interaction. In \appref{sec:app_SPA}, we show how the same result can be recovered using SPA-SALT. Note that in the single-emitter limit where $D_0 (\bvec{r}) \propto \delta(\bvec{r}-\bvec{r}^\prime)$, \eqref{eq:pump_thresh} simplifies to $d_\text{thresh} = V/Q$, where $V$ is the typical ``mode volume'' of the resonance, recovering an expression inversely proportional to the Purcell enhancement and the LDOS~\cite{LDOS_Laser, LDOS_Laser_2, LDOS_Laser_3}; we comment further on this regime in Sec.~\ref{sec:single}.

Once the pumping strength crosses the lasing threshold, the cavity-mode will start to lase with an amplitude $a_1$, as depicted in \figref{fig:schem}(b). Considering both the zeroth-order and the first-order correction in the small amplitude expansion of \eqref{eq:gain_amp}, we derive an approximate expression for the lasing intensity
\begin{equation} \label{eq:amp}
    |a_1|^2 \approx \frac{\Delta d}{ d_\text{thresh}}\frac{{\int_{\Omega} \varepsilon_c(\bvec{r}) |\mathbf{e}_{1}(\bvec{r})|^2 \d \Omega}\int_{\Omega} D_0(\bvec{r})|\mathbf{e}_{1}(\bvec{r})|^2 \,  \d \Omega} {e_c^{-2} \int_{\Omega} D_0(\bvec{r}) |\mathbf{e}_{1}(\bvec{r})|^4 \,  \d \Omega} \,,
\end{equation}
where we have assumed that we are operating in the weak pumping limit (just above threshold), where $ d \approx d_\text{thresh}$ and where $\Delta d=d-d_\text{thresh} \ll 1$ represents the pumping strength above threshold. For more details on this derivation, consult \appref{sec:app_B}. Note that evaluating \eqsref{eq:pump_thresh}{eq:amp} still requires one to solve a linear eigenproblem to determine the modes of the passive cavity, as in SPA-SALT. Nevertheless, assuming a high-$Q$ cavity $(Q\gtrsim
100$), we can eliminate this requirement by exciting the field in the cavity with an external source (e.g., a waveguide coupled to the cavity), since the field in the cavity will be approximately proportional to the passive cavity mode $\mathbf{e}_{1}$~\cite{phot_crys, Rasmus_FOM}. Although this eliminates the requirement of an eigensolve, it does not include information about the efficiency of coupling between the cavity mode and the output channel. In the next section, we will use TCMT to encode the out-coupling into a single, convenient FOM.


\subsection{Nonlinear FOM via temporal coupled mode theory}\label{sec:TCMT}

\begin{figure}[!h] 
    \centering
    \includegraphics[scale=1]{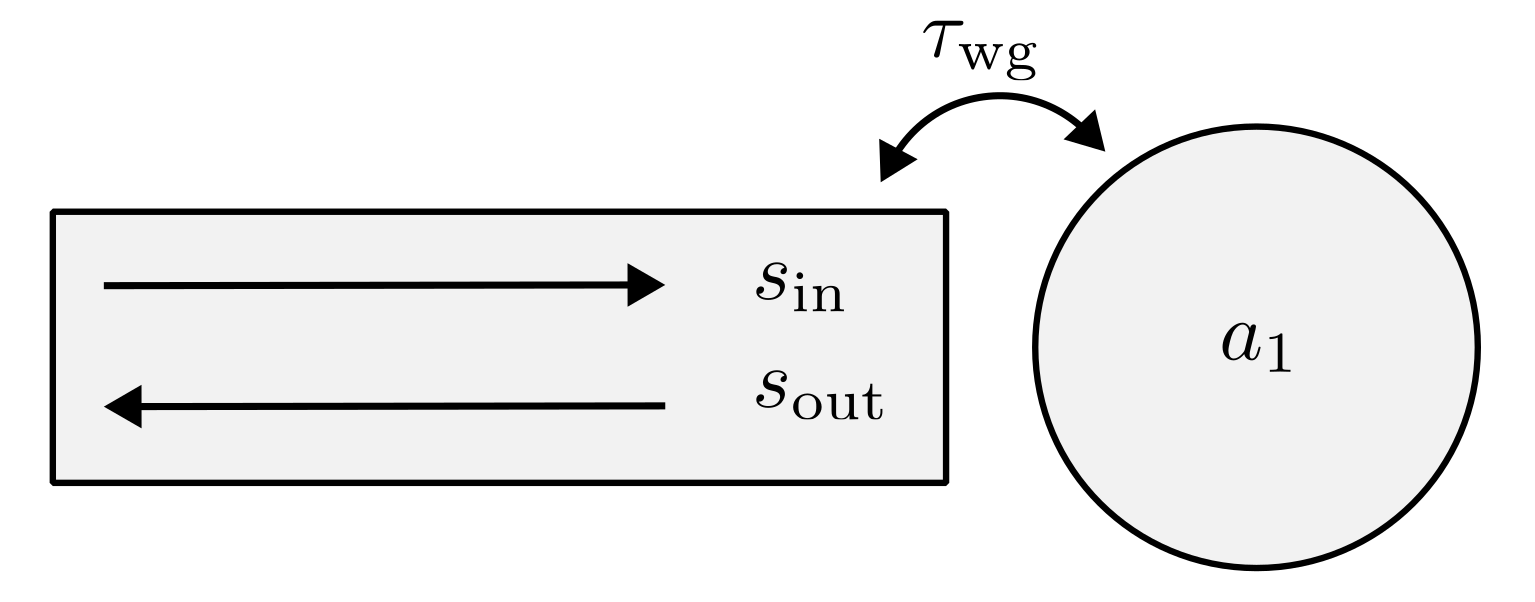} 
    \caption{Coupled optical system composed of a waveguide guiding an input field with amplitude $s_\text{in}$ and an output field with amplitude $s_\text{out}$, coupled to a cavity with an amplitude $a_1$ through the decay rate $\tau_\text{wg}$.} \label{fig:tcmt}
\end{figure}

To define a FOM that can be evaluated by solving a single linear system of equations and includes information about the coupling into the output channel, we use the expression for the amplitude in \eqref{eq:amp} in the context of TCMT.  For a system with a sufficiently strong resonance (e.g.~$Q \gtrsim 100$~\cite{phot_crys}) coupled to a discrete set of input/output ports, TCMT derives a minimal set of equations coupling the resonant amplitude with the amplitudes of the port modes, constrained by energy conservation and reciprocity to be determined purely in terms of the coupling rates $1/\tau$ (lifetimes $\tau$)~\cite{phot_crys}. Such a coupled optical system is illustrated in \figref{fig:tcmt}, where a cavity mode with amplitude $a_1$ (normalized so that $|a_1|^2$ is the energy in the cavity), \textit{total} decay rate of the resonance $1/\tau_1$, and quality factor $Q=\omega_1 \tau_1/2$, is coupled to a waveguide with an input amplitude $s_\text{in}$ and output amplitude $s_\text{out}$ (normalized so that $|s|^2$ is power) through a decay rate $1/\tau_\text{wg}$. Note that the \textit{total} decay rate 1/$\tau_1$ includes the decay rate $1/\tau_\text{wg}$ into the waveguide plus the decay rates into other loss channels such as absorption and/or radiation~\cite{phot_crys}.

\subsubsection{Emission problem}

In a direct formulation of the emission problem, the emitters in the gain medium will generate energy in the cavity mode, which then couples into the output waveguide mode. Using the TCMT equations, one can calculate the output power when there is no input waveguide mode ($s_{\text{in}} = 0$), as~\cite{phot_crys}
\begin{equation}
    P_\text{out} = |s_\text{out}|^2 = \frac{\omega_1}{Q_\text{wg}}|a_1|^2\,,
\end{equation}
where $Q_\text{wg} = \omega_1 \tau_\text{wg}/2$ is the quality factor for the coupling between cavity and waveguide. By substituting the expression for the amplitude in \eqref{eq:amp} and the expression for the threshold in \eqref{eq:pump_thresh}, the TCMT output power becomes
\begin{equation}\label{eq:out}
P_\text{out} \approx \Delta d \, \omega_1 \frac{Q}{Q_\text{wg}}\, \frac{\left(\int_{\Omega} D_0(\bvec{r})|\mathbf{e}_{1}(\bvec{r})|^2 \,  \d \Omega\right)^2} {e_c^{-2} \int_{\Omega} D_0(\bvec{r}) |\mathbf{e}_{1}(\bvec{r})|^4 \,  \d \Omega}.
\end{equation}
Note that, to evaluate this expression, one still needs to solve a single linear eigenvalue problem to determine the quality factors and mode profiles. 

\subsubsection{Reciprocal problem}

Instead, we now consider the reciprocal problem, where the input port excites the resonance (often a useful approach for ensembles of emitters~\cite{Yao_2022}). In this case, the TCMT equations imply a simple linear relationship between the input power and the resonant amplitude~\cite{phot_crys}
\begin{equation}\label{eq:energy}
    U = |a_1|^2 = \frac{2 \tau_1^2}{\tau_\text{wg}}|s_\text{in}|^2\ = \frac{1}{\omega_1} \frac{Q^2}{Q_\text{wg}}P_\text{in}\,.
\end{equation}
In the high-$Q$ system, the field from the reciprocal solve $\bvec{E}_{\mathrm{r}}$ is almost exactly equal to the cavity mode plus an $\bigO(1/\sqrt{Q})$ error~\cite{phot_crys}; thus, the energy $|a_1|^2$ can also be readily calculated by evaluating an integral of the reciprocal electric field in the cavity: $U = \int_{\Omega} \varepsilon_c(\bvec{r})|\bvec{E}_{\mathrm{r}}(\bvec{r})|^2  \d \Omega$. Similarly, with this argument, the reciprocal field $\bvec{E}_{\mathrm{r}}$ can also be used as a substitute for the modal field in \eqsref{eq:pump_thresh}{eq:out}.

\subsubsection{Nonlinear FOM for nanolaser optimization}\label{sec:FOM}

By solving the reciprocal problem in the weak pumping regime ($\Delta d \ll 1$), where we can model the pump power as being proportional to the pump strength ($P_\text{pump} \propto d \approx d_\text{thresh}$), we can define a nonlinear lasing FOM
\begin{equation}\label{eq:eff_nl}
     \frac{P_\text{out}}{P_\text{pump}} \propto \frac{\left( \int_{\Omega} D_0(\bvec{r})|\bvec{E}_{\mathrm{r}}(\bvec{r})|^2 \,  \d \Omega \right)^3} {\int_{\Omega} D_0(\bvec{r}) |\bvec{E}_{\mathrm{r}}(\bvec{r})|^4 \,  \d \Omega} = \text{FOM}.
\end{equation}
where we have used the expression of the output power in \eqref{eq:out}, and the lasing threshold expression in  \eqref{eq:pump_thresh}. We have left out proportionality factors and have substituted the energy integral with the expression derived in \eqref{eq:energy}. Although the ``efficiency'' ($P_\text{out}/P_\text{pump}$) on the left-hand-side of \eqref{eq:eff_nl} is $\leq 1$, eliminating the proportionality factors results in the FOM defined by the right-hand-side taking arbitrary (nonnegative) values. Notice that all quality factors simplify, and we are left with an expression that only depends on reciprocal field integrals, meaning that the FOM can be evaluated with a single reciprocal solve (excited \emph{from} the output waveguide). Moreover, this FOM $\sim |\bvec{E}_{\mathrm{r}}|^6 / |\bvec{E}_{\mathrm{r}}|^4 \sim |\bvec{E}_{\mathrm{r}}|^2$ is roughly proportional to the energy in the cavity ($\sim |\bvec{E}_{\mathrm{r}}|^2$) and thus to the $Q$~factor. This fact will ensure that maximizing the FOM yields high-$Q$ optimized cavities, making the perturbative single-pole approximation more and more accurate as the optimization progresses (numerical validation of this high-$Q$ approximation is provided in \appref{sec:app_val} for optimized nanolasers).

\subsubsection{LDOS and the single-emitter limit}\label{sec:single}

To show how the nonlinear/first-principles FOM compares to more conventional cavity-optimization approaches, we introduce a ``naive'' generalization of the LDOS, which \emph{heuristically} modifies the definition of the LDOS to account for out-coupling efficiency (through a reciprocal solve $\bvec{E}_{\mathrm{r}}$) and a distributed gain medium $D_0$
\begin{equation}\label{FoM:naive}
    \text{FOM}_\text{naive} = \int_{\Omega} D_0(\bvec{r}) |\bvec{E}_{\mathrm{r}}(\bvec{r})|^2 \d \Omega\,.
\end{equation}
This FOM targets the overlap of the electric-field intensity of the reciprocal field with the gain distribution. Like our nonlinear FOM, this naive FOM is also proportional to the electric-field intensity in the cavity ($\sim |\bvec{E}_{\mathrm{r}}|^2$) and thus to the $Q$~factor (all other things being equal, though this FOM is also affected by the coupling/extraction efficiency), which will also tend to yield high-$Q$ optimized cavities, with strong field in the gain medium. Similarly, it will favor a low lasing threshold via \eqref{eq:pump_thresh}. Indeed, we observe these properties in the inverse-design results of \secref{sec:results}.

Interestingly, in the single-emitter limit, where the gain distribution is modeled as being concentrated at a single point $\bvec{r^\prime}$, i.e.~$D_0(\bvec{r})\propto \delta (\bvec{r}-\bvec{r^\prime})$, the nonlinear FOM (\eqref{eq:eff_nl}) and the naive FOM (\eqref{FoM:naive}) become \emph{equivalent}, reducing to an LDOS-like FOM when solving the reciprocal problem~\cite{LDOS_reciprocal}: $\text{FOM} \propto |\bvec{E}_{\mathrm{r}}(\bvec{r^\prime})|^2$.  Whereas the usual definition of LDOS would correspond to the \emph{total} power expended by an orientation-averaged dipole source at $\bvec{r^\prime}$~\cite{OskooiJo13-sources}, this FOM corresponds (via reciprocity~\cite[App.~C]{LDOS_reciprocal}) to the dipole power emitted only into the output channel, a form of extraction efficiency.  Such a point-gain FOM is therefore reminiscent of earlier efforts~\cite{LDOS_opt_wang, LDOS_opt_liang, LDOS_bounds, albrechtsen_nanometer-scale_2022, LDOS_G} to optimize cavities by maximizing LDOS or extraction efficiency.

However, for extended active regions with a gain region whose diameter is not small compared to the wavelength ($\gtrsim \lambda$), the naive and  nonlinear objective functions are inequivalent, and we will show in \secref{sec:benchmark} that the resulting geometries and performance differ substantially, highlighting the importance of deriving the correct first-principles FOM rather than using heuristic targets. 

Another interesting situation is that of a single point-like gain region, such as a quantum dot, whose location randomly varies from one fabrication to the next.  In that case, one might wish to target the \emph{average} performance  (the expected value $E[\text{FOM}]$).  Since the point-like gain FOM is $\propto |\bvec{E}_{\mathrm{r}}(\bvec{r^\prime})|^2$, performing an ensemble average over gain locations $\bvec{r^\prime}$ with a probability distribution $\mathcal{P}(\bvec{r^\prime})$ immediately leads to $E[\text{FOM}] \propto \int \mathcal{P}(\bvec{r}) |\bvec{E}_{\mathrm{r}}(\bvec{r})|^2$, identical to \eqref{FoM:naive} except that $D_0$ is replaced by $\mathcal{P}$.

\subsection{Accounting for steady-state diffusion in the gain medium} \label{sec:diffusion_theory}

In semiconductor lasers with extended media, i.e., as opposed to isolated emitters such as quantum dots, it is paramount to account for diffusion of excited carriers. Thus, while the spatial profile of the lasing mode leads to a spatial variation of the local rate of stimulated emission and hence a spatially varying depletion of excited states, gain diffusion smoothens out the effective gain profile. As introduced in the complex-SALT (C-SALT) model~\cite{csalt}, this effect can be accounted for by modeling a bulk semiconductor gain medium in the free-carrier approximation, where carrier--carrier Coulomb interactions are neglected. In this approximation, one can include diffusion in \eqref{eq:gain_amp} by applying a set of linear operators
\begin{equation}\label{eq:gain_diff}
    \Delta \varepsilon_\Im(\bvec{r}) = \left(\mathbb{S}^{-1} + \mathbb{I} \, e_c^{-2}\, |\bvec{E}_L|^2 \right)^{-1} [D_0](\bvec{r})d\,
\end{equation}
where $\mathbb{I}$ is the identity operator, and the exponentially damped diffusion is encoded in the operator $\mathbb{S}^{-1}= \mathbb{I}+\nabla \cdot (R_\nabla^2(\bvec{r}) \nabla)$, where $R_\nabla (\bvec{r})$ is a diffusion lengthscale determined by the spatially- and design-dependent diffusion coefficient combined with the damping/recombination rate. In this notation, an operator is applied as $\mathbb{L}[b]$, where $\mathbb{L}$ is the operator and $b$ is a scalar field. As an example, computing $u = \mathbb{S}[b]$ corresponds to solving for the scalar field $u$ in the diffusion problem $\mathbb{S}^{-1}u=b$. Note that for small diffusion $R_\nabla \ll \lambda$   we recover the original expression in \eqref{eq:gain_SALT} because $\mathbb{S} \approx \mathbb{I}$. 

Similar to the derivations in \secref{sec:SPA}, we calculate the pumping strength needed to reach the lasing threshold by employing perturbation theory for small $|\bvec{E}_L|^2 \sim |a_1|^2$
\begin{equation}\label{eq:pump_thresh_diff}
    d_\text{thresh} = \frac{1}{Q} \frac{\int_{\Omega} \varepsilon_c(\bvec{r})|\mathbf{e}_{1}(\bvec{r})|^2\,  \d \Omega}{\int_{\Omega} \mathbb{S} [D_0] (\bvec{r}) |\mathbf{e}_{1}(\bvec{r})|^2\,  \d \Omega}\,.
\end{equation}
In this case, the lasing threshold can be decreased by overlapping the mode with the diffused gain profile $\mathbb{S} [D_0]$ in the cavity.

In a similar fashion to the derivations in \secref{sec:TCMT}, we use TCMT to calculate the nonlinear lasing FOM that accounts for diffusion
\begin{equation}\label{eq:eff_diff}
    \text{FOM} \propto  \frac{\left(\int_{\Omega} \mathbb{S} [D_0](\bvec{r}) |\bvec{E}_{\mathrm{r}}(\bvec{r})|^2 \,  \d \Omega\right)^3} {\int_{\Omega} \mathbb{S}\left[ |\bvec{E}_{\mathrm{r}}|^2\, \mathbb{S} [D_0] \right] (\bvec{r})|\bvec{E}_{\mathrm{r}}(\bvec{r})|^2 \,  \d \Omega}
\end{equation}
where the diffusion is applied to both the gain profile in the numerator, and to the product of the diffused gain profile $\mathbb{S}[D_0]$ and the electric-field intensity of the reciprocal field in the denominator. This highlights the fact that diffusion will not only affect the gain medium but will also diffuse depletion of the emitters in the gain region due to the hole-burning effect. For a more detailed derivation of the expressions in this section, refer to \appref{sec:app_C}.

\subsection{Topology optimization formulation}\label{sec:TopOpt}

In this section, we review the standard density-based TopOpt formulation \cite{topopt_dens, jensen_topology_2011} that we employ to describe and optimize our degrees of freedom, which is mostly independent of the choice of FOM.  Density-based TopOpt parameterizes the distribution of materials in space by introducing a design field,  the ``density'' $\rho(\bvec{r})\in[0,1]$, where the extremes $\rho = 0$ and $\rho = 1$ correspond to the two materials, while unphysical intermediate materials are also permitted temporarily as the design evolves. To regularize the optimization problem by introducing a weak sense of geometric lengthscale (ensuring convergence as the computational resolution is increased), one
applies a filtering and thresholding scheme to the design variables. The thresholding, which pushes the design towards a ``binarized'' structure in which intermediate material are excluded, is performed via a smoothed Heaviside projection~\cite{projection}
\begin{equation}\label{eq:thres}
    \hat{\rho} \equiv \Theta_{\beta, \eta}(\tilde{\rho}) = \frac{\tanh (\beta \cdot \eta)+\tanh (\beta
        \cdot(\tilde{\rho}-\eta))}{ \tanh (\beta \cdot \eta)+\tanh (\beta
        \cdot(1-\eta))}\,,
\end{equation}
where $\hat{\rho}$ is the physical field, $\Theta_{\beta, \eta}(\tilde{\rho})$ 
is the thresholding function, $\beta \in[1, \infty)$ and $\eta \in[0,1]$ are hyper-parameters that control the threshold sharpness and value
respectively.  The filtered design field $\tilde{\rho}$, which regularizes the problem by introducing a minimum lengthscale $r_{\text{f}}$, is obtained using a Helmholtz-based filter~\cite{lazarov_filter}
\begin{equation}\label{eq:filter}
-\left(\frac{r_{\text{f}}} {2\sqrt{3}}\right)^2\nabla^2\tilde{\rho}+\tilde{\rho}=\rho\,,
\end{equation}
where $r_{\text{f}}$ is a filter radius. The projection steepness $\beta$ is gradually increased as the optimization progresses: a small $\beta$ allows grayscale structures in which the topology can change smoothly, while a larger $\beta$ binarizes the structure but slows convergence (so the largest $\beta$ is only imposed at the end, when the structure is nearly converged)~\cite{ole_beta, HammondOs21}; the precise optimization details are given in \appref{sec:app_D}.

The filtering step regularizes the problem, but does not strictly exclude features that have arbitrarily small lengthscales.  In order to exclude such non-manufacturable results, we employ a standard technique to impose additional geometric minimum-lengthscale constraints ~\cite{lengthscale, meep} with a minimum feature size just above the single-pixel level ($\gtrsim 40$ nm) on both the solid and void regions, with the details in \appref{sec:app_D}.

The projected density $\hat{\rho}$ is mapped onto a physical material (and gain) by a straightforward interpolation.
In particular, the refractive index $n$ is given by a linear interpolation
\begin{equation}\label{eq:interp}
n =n_0+\hat{\rho}\left(n_1-n_0\right)\,, 
\end{equation}
where subscript $1$ denotes the material and $0$ the background. We also model a design-dependent extinction coefficient
\begin{equation}
\kappa = - ( \alpha + \alpha^\prime \hat{\rho} (1-\hat{\rho}))\,,
\end{equation}
where $\alpha$ is a coefficient that introduces artificial optical losses to broaden the frequency response of high-$Q$ optical resonances, thus facilitating the optimization problem~\cite{LDOS_opt_liang}; and $\alpha^\prime$ is an attenuation factor that discourages intermediate values of $\hat{\rho}$ by introducing optical losses. As the optimization progresses, $\alpha$ is gradually ``turned off'' to zero (eliminating the unphysical damping) while  $\alpha^\prime$ is gradually ``turned on'' (penalizing unphysical intermediate materials).  The gain medium is also dependent on $\hat{\rho}$, because we assume that gain is only present in the solid material. Furthermore, in \secref{sec:results}, we explore gain regions of varying diameters, described by a Gaussian profile around the center $\bvec{r}_0$ of the cavity, yielding the equation
\begin{equation}
    D_0 (\bvec{r}, \hat{\rho}) = \varepsilon_{\mathrm{c}(\bvec{r})}  \hat{\rho}(\bvec{r}) \, e^{- | \bvec{r} - \bvec{r}_0 |^2 / 2 \sigma_{\text{g}}^2 }\,,
\end{equation}
where $\sigma_{\text{g}}$ parametrizes the width of the Gaussian.  Future work may replace this ad-hoc gain profile with an explicit physical model of a pumping process, such as optical absorption, as discussed in \secref{sec:conclusion}.

\section{Computational Results}\label{sec:results}

In this section, we apply the theoretical framework reviewed and developed in the previous sections to optimize nanocavity-based laser devices. First, we compare our developed first-principles FOM with a heuristic generalization of the LDOS (\secref{sec:single}) through the inverse design 2d cavities (\secref{sec:benchmark}), demonstrating the increasing importance of a first-principles model as the diameter of the gain region increases. Then, we study how steady-state diffusion effects influence the optimized nanocavity designs (\secref{sec:diffusion_results}).  In all cases, the permittivity of the solid regions is $\varepsilon=12$ (characteristic of semiconductors) and of the void regions is $\varepsilon = 1$ (air). The gain medium material is modeled as a III--V semiconductor-like material with a direct bandgap. Lastly (\secref{sec:3D}), we consider the inverse design of more realistic 3d cavities in semiconductor-on-insulator system (SOI, $\varepsilon = 1.44^2$ for the substrate). (Detailed simulation and optimization parameters are given in \appref{sec:app_D}.)

In all cases, the laser is optimized for a wavelength of $1550$~nm, and we impose a minimum lengthscale~\cite{lengthscale} (\secref{sec:TopOpt}) just above the single-pixel level ($\gtrsim 40$ nm) on both solid and void regions.  The design region is $2.325\,\upmu\mathrm{m} \times 2.325\,\upmu\mathrm{m}$, connected to a single-mode output waveguide (width 500~nm), with PML absorbing boundary layers~\cite{taflove, BERENGER1994185} along with absorbing boundary conditions~\cite{jin}.  The 2d simulations employ the $H_z$ (out-of-plane $\bvec{H}$, in-plane $\bvec{E}$) polarization, whereas the 3d simulations use the analogous mostly-$H_z$ waveguide mode.  All simulations employ a finite-element discretization whose details are given in \appref{sec:app_D}, with resolution $\approx 40$~nm in the design region.  Antisymmetric mirror-symmetry boundary conditions (Neumann for $H_z$, equivalent to a perfect electric conductor) are imposed to halve the computational region (bisecting the waveguide). Although such a non-convex optimization problem may, in principle, have non-symmetric local optima (and TopOpt often exhibits many local optima with similar performance~\cite{ChenCh24, dalklint2024performanceboundstopologyoptimization}), we found that symmetric optima were often found even if we did not impose symmetry, and imposing mirror symmetry reduces computational costs.  Except where otherwise noted, all optimizations began with a homogeneous $\rho = 0.5$ starting structure.

\subsection{2d: Nonlinear vs.~naive FOM}\label{sec:benchmark}

\begin{figure*}[!t] 
    \centering
    \includegraphics[scale=1]{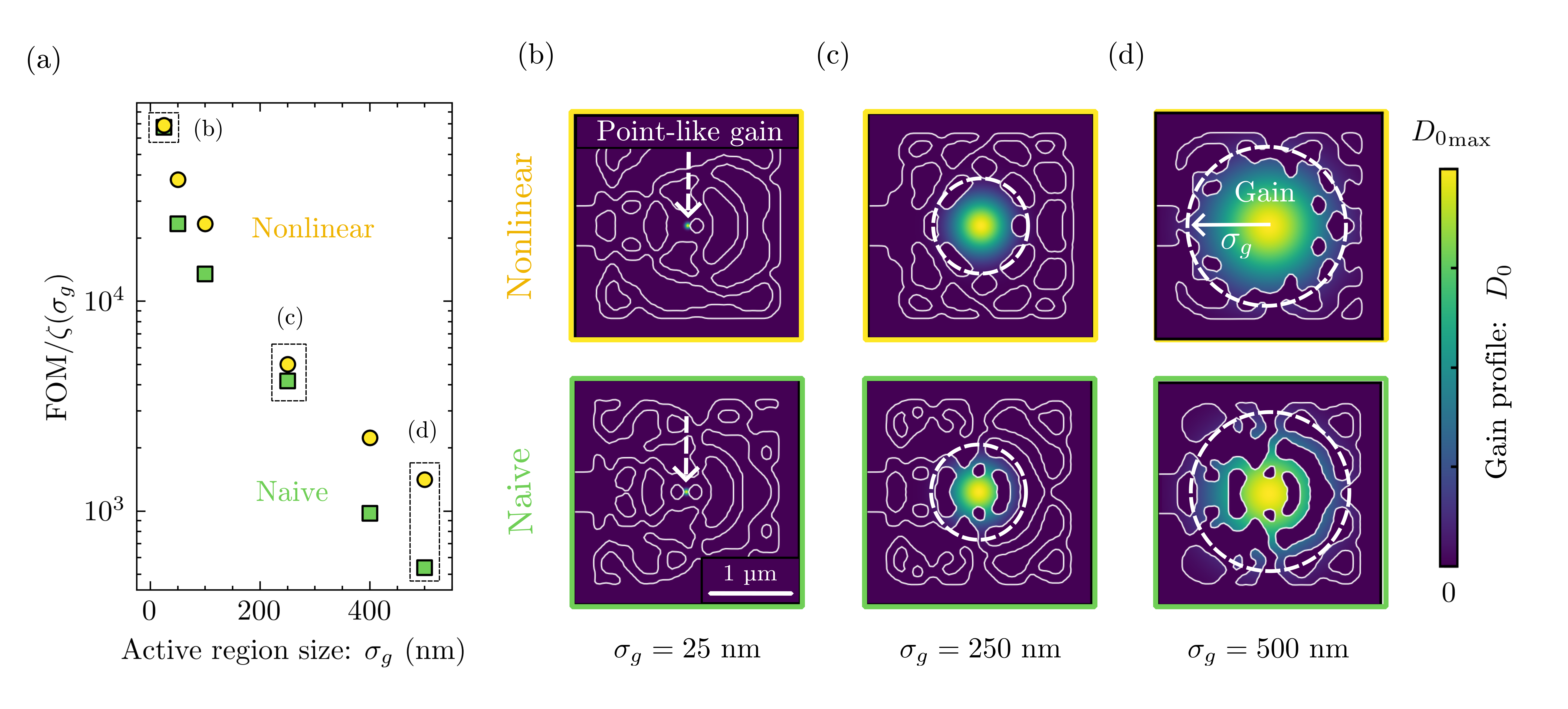} 
    \caption{Performance of inverse designed devices for increasing active region sizes $\sigma_{\text{g}}$. (a) Evaluation of the nonlinear FOM normalized by the gain region size-dependent factor $\zeta(\sigma_g)=\max\left\{|\mathbf{E}_\text{in}|^2\right\}\left(\int_\Omega D^*_0(\sigma_g) \d \Omega\right)^2$, for designs optimized for the naive FOM (green) and nonlinear FOM (yellow). The gain profile $D_0$ is shown in (b), (c), and (d) in max-normalized units for increasing active region sizes. The standard deviation $\sigma_{\text{g}}$ of the gain profile is marked in a dashed white line. In the gain profile for the nonlinear device in (d), the active region size $\sigma_{\text{g}}$ is explicitly shown for illustration purposes.}
    \label{fig:nonlinear}
\end{figure*}

To showcase the difference of optimizing for a naive generalization of the LDOS (\eqref{FoM:naive}, \secref{sec:single}) or for the first-principles nonlinear FOM (\eqref{eq:eff_nl}, \secref{sec:FOM}), in \figref{fig:nonlinear} we compare the performance of optimized nanolasers for different gain region sizes: $\sigma_{\text{g}} \in \lbrace 25\text{ nm},  50\text{ nm}, 100\text{ nm}, 250\text{ nm}, 400 \text{ nm}$, $500\text{ nm}\rbrace$. Half of the cavities are optimized for the naive FOM (green) , while the other half are optimized for the derived nonlinear FOM (yellow). To make the FOM easier to interpret, reproduce, and compare across nanolaser platforms, we normalize the nonlinear FOM by dividing by $\zeta(\sigma_g)=\max\left\{|\mathbf{E}_\text{in}|^2\right\}\left(\int_\Omega D^*_0(\sigma_g) \d \Omega\right)^2$, where $\mathbf{E}_{\mathrm{in}}$ is the input‐waveguide electric field and $D^*_0=e^{- | \bvec{r} - \bvec{r}_0 |^2 / 2 \sigma_{\text{g}}^2}$ is the Gaussian gain profile with the structure ($\varepsilon_{\mathrm{r}} \hat{\rho}$) dependence removed. Note that the results have been cross checked to ensure that indeed a design optimized for a certain domain size is the best for that domain size.

As pointed out in \secref{sec:single}, in the single-emitter limit where the active region is point-like, the naive and nonlinear FOM become equivalent, and in consequence we observe in \figref{fig:nonlinear}(b) that the optimized cavities become nearly equivalent  for small $\sigma_{\text{g}} \ll \lambda$.  In this single-emitter limit, the cavity geometry exhibits a central bowtie-like feature---a structure commonly arising in LDOS optimization~\cite{LDOS_opt_wang, LDOS_opt_liang, LDOS_bounds, albrechtsen_nanometer-scale_2022, LDOS_G}---which is known to enhance light–-matter interaction at a single point~\cite{choi_self-similar_2017, albrechtsen_two_2022, YaoVe23} by concentrating the in-plane electric field at the emitter location thanks to a field singularity~\cite{sing, bowtie_laser} at any sharp tip.  (Here, the bowtie sharpness is limited by our $\gtrsim 40$ nm lengthscale constraints.) Due to the strong field concentration at the cavity-tip, the normalized FOM/$\zeta(\sigma_g)$, has a larger value compared to the cavities that were optimized for larger gain-region sizes.

\begin{figure}[!h] 
    \centering
    \includegraphics[scale=1]{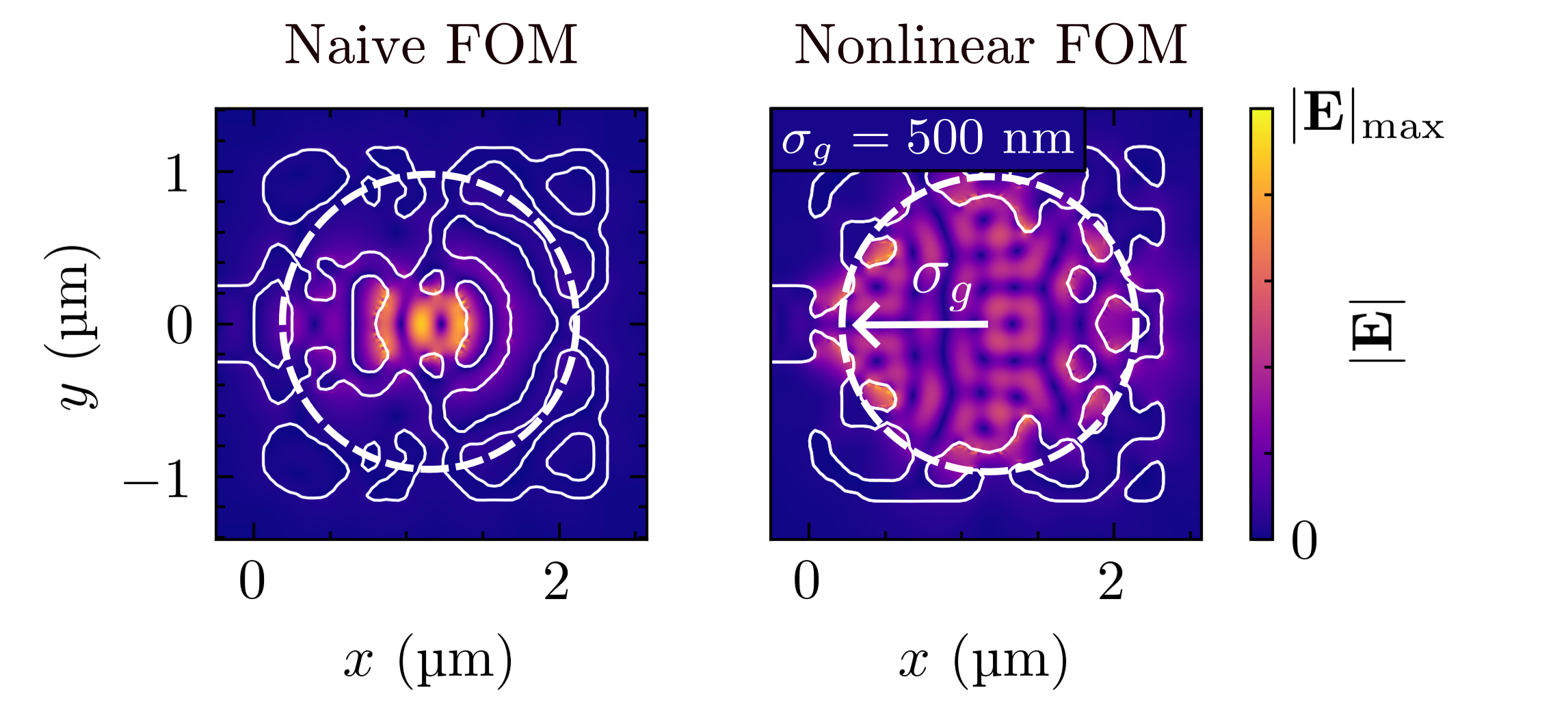} 
    \caption{Electric-field norm $|\bvec{E}(\bvec{r})|$ in devices optimized for a gain medium of standard deviation $\sigma_{\text{g}} = 500$~nm. The fields are expressed in max-normalized units.}
    \label{fig:benchmark_fields}
\end{figure}

As the active region size increases, the system transitions beyond the single-emitter regime, and the cavity designs optimized for the naive and nonlinear FOMs begin to differ significantly as the gain diameter approaches the wavelength ($\sigma_{\text{g}} \gtrsim \lambda/n_1 \sim 450$~nm). In both cases, the spatial averaging of the emission means that arbitrarily sharp tips are no longer beneficial (because $\int | \bvec{E} |^2$ is finite)~\cite{YaoVe23}. For our largest gain diameter [$\sigma_{\text{g}} = 500$~nm, \figref{fig:nonlinear}(d)], the structure optimized using the nonlinear FOM achieves a performance $\approx 3 \times$ greater than the structure optimized with the naive FOM. We attribute this to the nonlinear ``hole-burning'' effect (\secref{sec:SALT}) captured in the denominator of the nonlinear FOM (\eqref{eq:eff_nl}) but not present in the naive FOM (\eqref{FoM:naive}), which penalizes strong spatial field localization. As illustrated in \autoref{fig:benchmark_fields}, the electric-field distribution shows that the cavity optimized for the naive FOM concentrates the field at its center, whereas the cavity optimized for the nonlinear FOM distributes the field more homogeneously around the active region $D_0$, reducing hole burning and leading to a higher nonlinear FOM. Moreover, as the active region size becomes larger, the normalized FOM ($\text{FOM}/\zeta(\sigma_g)$) decreases [\figref{fig:nonlinear}(a)], and transitions from the sharp-tip large-enhancement regime to distributing the electric field within the active material, which results in the normalized FOM becoming roughly proportional to $Q$ for larger active region sizes (\secref{sec:FOM}). Note that the non-normalized FOM increases by several orders of magnitude as the active‐region volume grows, simply reflecting the larger gain medium emitting light into the lasing mode. We emphasize that field singularities at sharp tips play a negligible role in most of the designs. As discussed above, they are only favored when the gain region is effectively point-like and the two FOMs become LDOS-like, whereas they are explicitly penalized in the nonlinear FOM for an active region with size comparable to the wavelength of light. This suppression of singular features contributes to the intrinsic robustness of the optimized cavities, as the performance does not rely on the minimum lengthscale of the geometry.

Finally, we validate \textit{a posteriori} one of the main assumptions in the first-principles modeling based on the SPA-SALT (\secref{sec:SPA}): that the optimized cavities are indeed high-$Q$ cavities. By means of a post-processing eigensolve on the optimized the cavities, we find that the $Q$s of the lasing modes are in the range $Q\in[350,1000]\gtrsim
100$, which validates the SPA-SALT and TCMT approximations~\cite{cerjan_2016, phot_crys} as shown in \appref{sec:app_val}. Not only that, as pointed out in \secref{sec:intro}, there is a trade-off between designing a high-$Q$ cavity and a cavity that couples efficiently to an output port. By computing the power fraction in the waveguide relative to the total lasing-mode power in the post-processing eigensolve, we evaluate the lasing mode's optical extraction efficiency in the topology-optimized cavities. As expected from the definition of the nonlinear FOM (\eqref{eq:eff_nl}), which directly incorporates out-coupling effects, we find values of the optical extraction efficiency $\geq0.9$ for most of the designs, highlighting the ability of the FOM to yield high-$Q$ cavities with high optical extraction efficiencies. We finally note that we have also assessed fabrication robustness by evaluating how systematic over– and under–etching would affect performance using the nonlinear FOM for the device with $\sigma=500$ nm. Following~\cite{eta}, a change in threshold value $\Delta\eta=0.1$ corresponds to a dimensional variation of $\Delta l\approx 10$\,nm, yielding a FOM scaled by a factor $\approx \times 0.20$ when evaluated at the new (shifted) resonance frequency as determined by an eigensolver. Moreover, to study how realistic random roughness (e.g., surface roughness) affects the design, we have considered a random variation of $\Delta\eta=0.1$ with a spatial correlation of 10 nm, and see that the FOM is scaled by a factor $\approx \times 0.55$ for such perturbations. This provides a quantitative estimate of expected performance tolerance to nanoscale variations for the optimized high-$Q$ nanolaser designs.

\subsection{2d: Gain diffusion}\label{sec:diffusion_results}

\begin{figure}[!h] 
    \centering
    \includegraphics[scale=1]{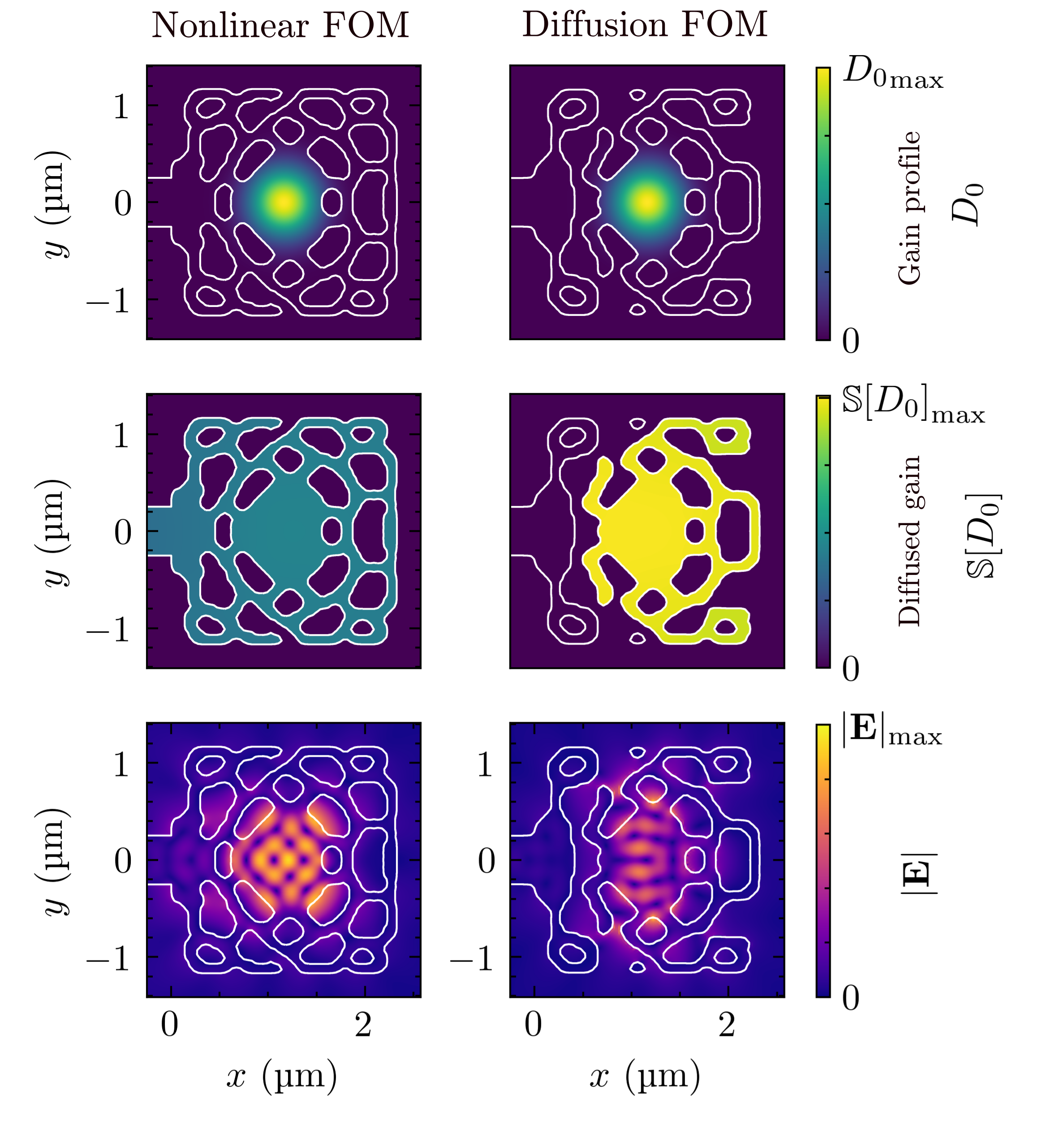} 
    \caption{Spatial profile of the gain medium $D_0(\bvec{r})$ with a standard deviation $\sigma_{\text{g}}=250$~nm, the diffused gain medium $\mathbb{S}[D_0](\bvec{r})$, for a diffusion length of $R_0=5\,\upmu\text{m}$, and the electric-field norm $|\bvec{E}(\bvec{r})|$ of the cavity mode for the inverse designed cavity not accounting (left) and accounting (right) for diffusion. All fields are expressed in max-normalized units.}
    \label{fig:diffusion}
\end{figure}

Gain diffusion is important in extended gain media, such as bulk materials and quantum wells, when the extent of the gain region becomes comparable to or larger than the diffusion length. In such cases, the diffusion counteracts the effect of spatial-hole burning. To understand the impact of gain diffusion on the optimization results we use a diffusion model based on the C-SALT framework \cite{csalt}(\secref{sec:diffusion_theory}) to analyze the performance of the cavity previously optimized for a gain medium of size $\sigma_{\text{g}} = 250$ nm (\figref{fig:nonlinear}), where diffusion was neglected. Using this design as an initial guess, we re-optimize the cavity in \figref{fig:diffusion} using the diffusion FOM (\eqref{eq:eff_diff}), assuming a diffusion length of $R_0 = 5\,\upmu\text{m}$ comparable to typical electron and hole diffusion lengths in semiconductor materials such as Si~\cite{Diff_Si, semicon_params}, InP~\cite{semicon_params}, (In,Ga)N~\cite{PhysRevApplied.15.054015}, InAs~\cite{semicon_params}, GaAs~\cite{semicon_params}, and InGaAs~\cite{PhysRevB.70.205311}. Note that as the diffusion lengths decrease, the system approaches the non-diffusive limit ($R^2_\nabla \ll 1$, \eqref{eq:gain_diff}), and the optimization results will converge toward those presented in \secref{sec:benchmark}.

Evaluating the diffusion-corrected performance of the device optimized without considering diffusion ($\text{FOM}/\zeta\approx1119$,``Nonlinear FOM'' in \figref{fig:diffusion}) reveals a significant 2-fold drop in FOM compared to the diffusion-optimized structure ($\text{FOM}/\zeta\approx2118$, ``Diffusion FOM'' in \figref{fig:diffusion}), underscoring the importance of including diffusion effects in the optimization process. This reduction in FOM can be attributed to the diffusion of carriers within the active region, as shown in the diffused gain profile $\mathbb{S}[D_0]$, which reduces the field overlap in a structure designed for the nominal Gaussian gain profile. However, by optimizing for the FOM that incorporates diffusion, the cavity design is able to compensate for this effect: both changing the field profile to match the modified gain distribution, and by changing the structure to control the diffusion itself. In particular, as shown in \figref{fig:diffusion}, the optimizer disconnects the cavity from both the waveguide and low-field regions, while also removing material from areas with weak electric field. This confines the emitters or carriers to the high-field cavity region and improves the overlap between the diffused gain region $\mathbb{S}[D_0]$ and the electric field distribution (which is also more spread out than in the non-diffusion design).

\subsection{Cavity design of a 3d physics model} \label{sec:3D}

In this section, we demonstrate that our approach is computationally efficient enough to attack fully three-dimensional systems, since it is no more expensive than traditional linear inverse design. In particular, we consider an example nanolaser cavity designed for typical semiconductor-on-insulator platforms, in which a 220~nm semiconductor slab is patterned on top of a silica ($n=1.44$) substrate, with air above.   The design region is otherwise similar to the 2d geometries, consisting of a 2d pattern and waveguide extruded with constant cross sections in the semiconductor layer, where voids in this layer are filled with air and the active material takes up the entire thickness (extrusion length) of the slab and is embedded within the semiconductor structure. PML+absorbing boundaries are used above and below the slab, separated from the semiconductor layer by $720$~nm.  Additional implementation details are provided in \appref{sec:app_D}.

\begin{figure*}[t!] 
    \centering
    \includegraphics[scale=1]{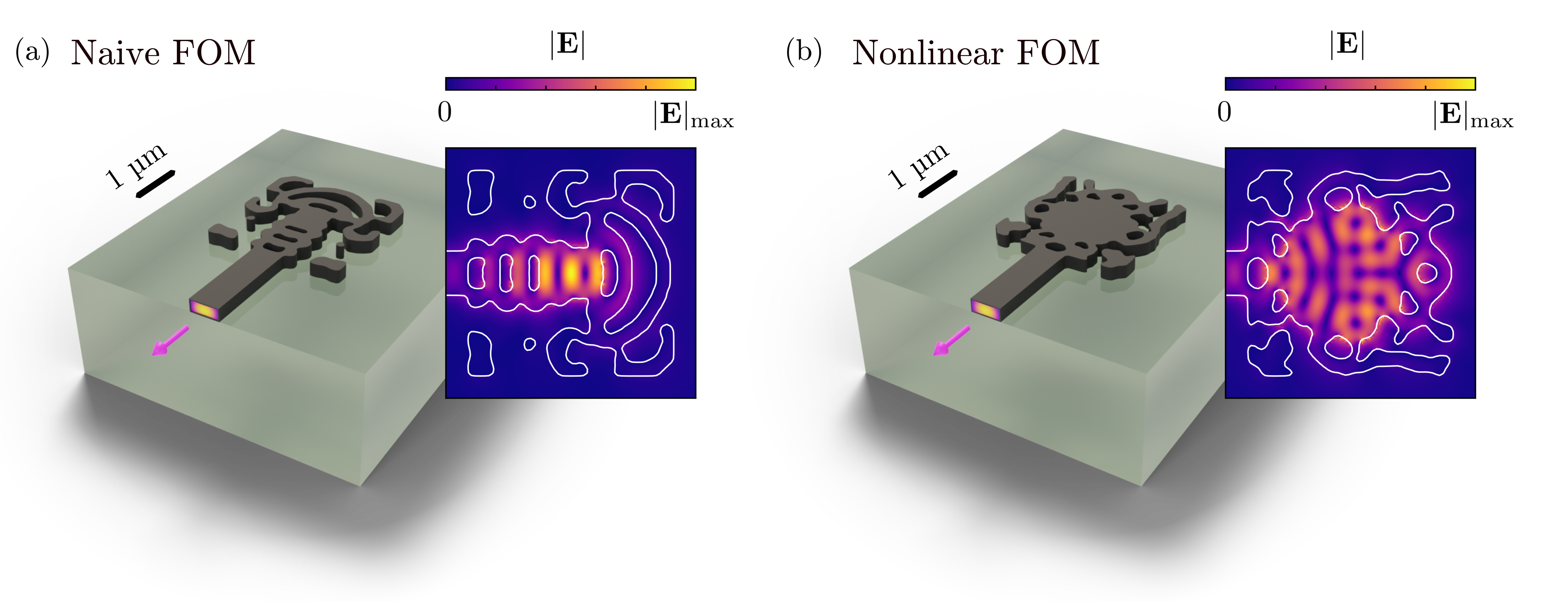} 
    \caption{Inverse designed devices in three dimensions. The devices lase into the cavity mode (inset plot) and out-couple to the fundamental TM mode of the output waveguide (pink arrow). The electric-field norm is given in max-normalized units. (a) Device optimized for the naive FOM. (b) Device optimized for the nonlinear FOM.}
    \label{fig:3D}
\end{figure*}

Using this 3d model, we again optimize nanolaser cavities with a gain medium size of $\sigma_{\text{g}}=500$~nm for the naive and nonlinear FOMs (\figref{fig:3D}), with no gain diffusion. Similar to the 2d case, the device optimized for the naive FOM creates a strong field concentration close to the cavity center, while the device optimized for the nonlinear FOM spreads the field more evenly over the active material region to counteract hole-burning effects. However, and in contrast to the 2d optimization results (\secref{sec:benchmark}), in 3d there is a more modest increase of the FOM when targeting nonlinear ($\text{FOM}/\zeta\approx148$) vs the naive FOM ($\text{FOM}/\zeta\approx90$) ---there is an increase of a factor of $\approx 1.6$ (instead of the factor of 3 in \secref{sec:benchmark}). We attribute this result qualitatively to the fact that localizing a high-$Q$ resonance is more difficult in a 3d slab system than in 2d, because of the need to minimize out-of-plane radiation. That is, in 3d, more design freedom is devoted to increasing $Q$ than in 2d, leaving less design freedom to optimize the field profile for hole-burning effects.  Nevertheless, the increase in the FOM is still a substantial improvement, with no additional computational cost \text{over a naive approach}. Post-processing the 3d results through an eigensolve again validates the high-$Q$ assumption for SPA-SALT and TCMT analysis, with values of $Q \approx 150$, while the optical extraction efficiency is moderately high $\geq 0.75$,  lower than in 2d (\secref{sec:benchmark}) presumably due to out-of-plane radiation for this tiny ($\approx 1.5\lambda \times 1.5\lambda)$ cavity.  We expect that higher $Q$ and efficiency should occur for larger design regions, similar to previous nanocavity designs.

\section{Conclusion}\label{sec:conclusion}

In this paper, we have shown that by careful analysis of a lasing model, exploiting perturbative analysis valid in high-$Q$ cavities, we can obtain an optimization figure of merit (FOM) for the input-to-output efficiency of the laser that that incorporates a wide range of important phenomena---such as resonant enhancement, spatial hole-burning, and gain diffusion---with little or no additional computational cost compared to simpler FOMs such as the LDOS.  Only a single \emph{linear} ``reciprocal'' Maxwell solve is required to evaluate our FOM, plus two scalar damped-diffusion solves if gain diffusion is included.   The results, in both two and three dimensions, show substantial benefits to such a first-principles approach compared to heuristic field-intensity/LDOS-like figures of merit, especially for a finite-diameter gain region.

While we believe that our new FOM should be immediately valuable to inverse design of nanolasers, it also offers a starting point for many future refinements.   An obvious next step would be to replace our \text{ad-hoc} gain profile $D_0$ with a full first-principles model of the pumping process itself, since that process is likely to also depend somewhat on the geometry.  For example, optical pumping can be modeled as a second Maxwell solve at the pump wavelength (e.g.~for vertical illumination), with $D_0$ being computed from the pump absorption profile~\cite{csalt, I-SALT, siegman86}.  This is conceptually similar to the coupled pump--emission solves that have been applied to inverse design of Raman scattering~\cite{ChristiansenMi20,PanCh21,YaoVe23}, scintillation~\cite{Roques_Carmes_2022}, and nonlinear harmonic generation~\cite{LinLi16,Mann2023}. Moreover, one could include a geometry-dependent effect on the pump power to reach transparency (\secref{sec:SALT}), as well as other effects such as absorption in the non-pumped region. Alternatively, modeling electrical pumping would involve solving a carrier-transport equation for $D_0$~\cite{chow1999semiconductor,OrtwinHess}, again coupled to the reciprocal lasing solve. Also, for the ultra-small active regions considered in nanolasers, pump blocking due to bandfilling effects become an important effect~\cite{saldutti2023thresholdsemiconductornanolasers}. Even more complicated pumping models might involve nonlinear interactions between the pumping and emission processes, but since the emission should still simplify via a resonance approximation, we are hopeful that such effects will still be a tractable nonlinear solve (e.g.~by a few Newton steps~\cite{EsterhazyLiu14} for pre-selected poles), which would be differentiable by implicit adjoint methods~\cite{Griewank2008}. In principle, optimization could also be applied to the full SALT equations, with the perturbative approach providing an initial guess that may reduce the difficulty of pole tracking---such a full-SALT optimization may become necessary if one wants to drive the system into the multi-mode regime. (Even without optimization, a SALT solve could be useful to quantify the accuracy of the SPA-SALT and TCMT approximations, although based on previous work and the numerical validation provided in \appref{sec:app_val}, we expect the errors in such perturbative methods to be small for high-$Q$ single-mode cavities like the results in this paper.) Another effect that might be interesting to include in an optimization FOM is the lasing linewidth, for which accurate first-principles formulas in terms of the lasing mode have recently been derived that include the full effect of nano-patterned geometries~\cite{pick}, and which should be equally applicable to our reciprocal-solve estimate of the lasing mode in the high-$Q$ regime. It has already been experimentally demonstrated that the distribution of active material in a nanostructured cavity can be used to significantly decrease the linewidth of a nanolaser~\cite{fano}.

As we commented in \secref{sec:single}, the ``naive'' FOM~(\ref{FoM:naive}) that we employ for comparison purposes is actually a \emph{correct} FOM for situations in which a \emph{single} emitter is randomly distributed with probability density $\sim D_0$.   This situation arises experimentally in systems employing quantum dots~\cite{pawel, QD}, color centers~\cite{Lukin_color, Stas_2022, Sipahigil_2016}, and similar isolated point-like emitters, whose placement is difficult to precisely control relative to the fabricated pattern.  Despite its simplicity, \eqref{FoM:naive} still encapsulates both the perturbative lasing physics and the out-coupling efficiency for such applications.

\startappendices

\section{Comparison with SPA-SALT}\label{sec:app_SPA}
\subsection{Single mode lasing threshold}
In this appendix, we derive the connection between the single pole threshold approximation in \secref{sec:SPA} and SPA-SALT. In SPA-SALT notation, Let a passive cavity resonance $\mathbf{e}_{1}(\bvec{r})$ satisfy an equation similar to \eqref{eq:SPA_SALT}, but without the gain medium,
\begin{equation}
{\left[(\nabla \times 
 \nabla \times ) -\varepsilon_c(\bvec{r})  \left(\frac{\tilde{\omega}_{1}^2}{c^2}\right)\right] \mathbf{e}_{1}(\bvec{r})}=0. \label{eq:passCav}
\end{equation}
Here, $\tilde{\omega}_{1}$ is the complex frequency of the passive cavity resonance. Under the single-pole approximation, the lasing mode $\bvec{E}_L(\bvec{r}) \approx a_1 \mathbf{e}_1(\bvec{r}) / \sqrt{\int_{\Omega} \varepsilon_c(\bvec{r}) |\mathbf{e}_{\mathrm{1}}(\bvec{r})|^2 \d \Omega} $, where $a_1$ is the complex amplitude of the field for the first lasing mode. Thus, multiplying \eqref{eq:passCav} by $a_1$ and equating \eqreftwo{eq:SPA_SALT}{eq:passCav}, 
\begin{equation}
    \varepsilon_c(\bvec{r}) \left(\frac{\tilde{\omega}_1^2}{c^2}\right) \mathbf{e}_1(\bvec{r}) = \Big(\varepsilon_c(\bvec{r})-i \Delta \varepsilon_\Im (\bvec{r})\Big) \left(\frac{\omega_L^2}{c^2}\right)\mathbf{e}_1(\bvec{r}).
\end{equation}
Rearranging,
\begin{equation}
    \varepsilon_c(\bvec{r}) \left(\frac{\tilde{\omega}_1^2 - \omega_L^2}{\omega_L^2} \right) \mathbf{e}_1(\bvec{r}) = -i \Delta \varepsilon_\Im (\bvec{r}) \mathbf{e}_1(\bvec{r}),
\end{equation}
The SPA-SALT lasing threshold is found by assuming that there is no hole-burning term in the gain medium and allowing the lasing frequency to be different from the gain's central frequency, such that 
\begin{equation}\label{eq:a3}
    \Delta \varepsilon_\Im (\bvec{r}) =  \left(\frac{\gamma_{\perp}}{\gamma_{\perp} - i(\omega_L-\omega_a)} \right) \frac{D_0(\bvec{r}) d}{1+ \Gamma_1 e_c^{-2} |a_1|^2 \left|\mathbf{e}_1(\bvec{r})\right|^2/ \int_{\Omega} \varepsilon_c(\bvec{r}) |\mathbf{e}_{\mathrm{1}}(\bvec{r})|^2 \d \Omega},
\end{equation}
becomes
\begin{equation}\label{eq:a4}
    \Delta \varepsilon_\Im^{(\textrm{thresh})} (\bvec{r}) =  \left(\frac{\gamma_{\perp}}{\gamma_{\perp} - i(\omega_L-\omega_a)} \right) D_0(\bvec{r}) d,
\end{equation}
where as a reminder $\Gamma_1= \gamma_\perp^2 / \left[\gamma_\perp^2+(\omega_L-\omega_a)^2\right]$. Altogether, at the first lasing threshold, this yields
\begin{equation}
    \varepsilon_c(\bvec{r}) \left(\frac{\tilde{\omega}_1^2 - \omega_L^2}{\omega_L^2} \right) \mathbf{e}_1(\bvec{r}) = -i \left(\frac{\gamma_{\perp}}{\gamma_{\perp} - i(\omega_L-\omega_a)} \right) d_{\textrm{thresh}} D_0(\bvec{r}) \mathbf{e}_1(\bvec{r}).
\end{equation}
The SPA-SALT threshold prediction~\cite{Ge_2010,cerjan_2016} is then found by integrating both sides with respect to $\mathbf{e}_1(\bvec{r})$ as SALT generally uses the biorthogonal normalization of the system's lasing modes and resonances. We can instead recover \eqref{eq:pump_thresh} by restoring the assumption that $\omega_L = \omega_a$, $\tilde{\omega}_1 = \omega_L - i\gamma_1$, and expanding $\left(\tilde{\omega}_L^2 - \omega_L^2\right)/\omega_L^2$ to first order in $\gamma_1$, yielding
\begin{equation}
    \varepsilon_c(\bvec{r}) \left(\frac{-2i \gamma_1}{\omega_L} \right) \mathbf{e}_1(\bvec{r}) = -i d_{\textrm{thresh}} D_0(\bvec{r}) \mathbf{e}_1(\bvec{r}),
\end{equation}
which is then integrated by $\mathbf{e}_1^*(\bvec{r})$,
\begin{equation}
    d_{\textrm{thresh}} = \left(\frac{2\gamma_1}{\omega_L} \right) \left( \frac{\int_\Omega \mathbf{e}_1^*(\mathbf{r}) \varepsilon_c(\mathbf{r}) \mathbf{e}_1(\mathbf{r}) d\Omega}{\int_\Omega \mathbf{e}_1^*(\mathbf{r}) D_0(\mathbf{r}) \mathbf{e}_1(\mathbf{r}) d\Omega} \right).
\end{equation}

\subsection{Interacting threshold for a second lasing mode}

To estimate the effect of spatial hole-burning on the thresholds of other potential lasing modes, we again approximate the spatial profile of each potential mode to be that of a different cavity resonance, $\mathbf{e}_m(\mathbf{r})$. We note that while the approximation of treating the first lasing mode as having the same spatial profile of a high-$Q$ resonance is quite good when the passive cavity resonance frequency is well-aligned with the central frequency of the gain (as we do here, see \appref{sec:app_val}), this same approximation on other lasing modes may be worse, especially if the resonance frequencies are outside of the gain window and thus experience significant line-pulling. Nevertheless, for the systems we consider here, which are tailored to support a single high-$Q$ passive cavity resonance, this approximation is still sufficient to estimate the range of single-mode behavior.

For any second lasing mode to reach threshold, such that the total lasing field can be decomposed into two different modes with distinct lasing frequencies $\mathbf{E}_L(\mathbf{r}; t) \approx a_1 \mathbf{e}_1(\mathbf{r})e^{-i \omega_L t} + a_2 \mathbf{e}_2(\mathbf{r})e^{-i \omega_{L_2} t}$, we seek a non-trivial (i.e., for $|a_2| \ll 1$, but $|a_2| \ne 0$) solution to
\begin{equation}
    \varepsilon_c(\bvec{r}) \left(\frac{\tilde{\omega}_2^2 - \omega_{L_2}^2}{\omega_{L_2}^2} \right) \mathbf{e}_2(\bvec{r}) = -i \left(\frac{\gamma_{\perp}}{\gamma_{\perp} - i(\omega_{L_2}-\omega_a)} \right) \left(\frac{D_0(\bvec{r}) d^{(2)}_{\textrm{int thr}}}{1+ \Gamma_1 e_c^{-2} \left|a_1\right|^2 \left|\mathbf{e}_1(\bvec{r})\right|^2 / \int_{\Omega} \varepsilon_c(\bvec{r}) |\mathbf{e}_{\mathrm{1}}(\bvec{r})|^2 \d \Omega} \right)  \mathbf{e}_2(\bvec{r}).
\end{equation}
Note that if $|a_1|=0$, the non-interacting threshold for this other mode is given by
\begin{equation}
    d^{(2)}_{\textrm{thr}} = i \left(\frac{\tilde{\omega}_2^2 - \omega_{L_2}^2}{\omega_{L_2}^2} \right) \left(\frac{\gamma_{\perp} - i(\omega_{L_2}-\omega_a)}{\gamma_{\perp}} \right) \left( \frac{\int_\Omega \mathbf{e}_2^*(\mathbf{r}) \varepsilon_c(\mathbf{r}) \mathbf{e}_2(\mathbf{r}) d\Omega}{\int_\Omega \mathbf{e}_2^*(\mathbf{r}) D_0(\mathbf{r}) \mathbf{e}_2(\mathbf{r}) d\Omega} \right),
\end{equation}
thus
\begin{equation}
    d^{(2)}_{\textrm{int thr}} = d^{(2)}_{\textrm{thr}} \left[1 + \Gamma_1 e_c^{-2} |a_1|^2 \frac{\int_\Omega \varepsilon_c(\mathbf{r}) |\mathbf{e}_2(\mathbf{r})|^2 |\mathbf{e}_1(\mathbf{r})|^2 d\Omega}{\left(\int_\Omega \varepsilon_c(\mathbf{r}) |\mathbf{e}_2(\mathbf{r})|^2 d\Omega\right)\left(\int_\Omega \varepsilon_c(\mathbf{r}) |\mathbf{e}_1(\mathbf{r})|^2 d\Omega \right)} \right]. \label{eq:a11}
\end{equation}
To completely specify the interacting threshold of a second lasing mode within the SPA-SALT approximation (though noting that here we are using the Hermitian normalization and approximate orthogonality), we can solve for $|a_1|$ at arbitrary $d$ in the single mode regime up to and including the second interacting threshold using Eqs.~\ref{eq:a3} and \ref{eq:a4} as
\begin{equation} 
    \frac{d}{d_{\textrm{thresh}}} - 1 = \Gamma_1 e_c^{-2}|a_1|^2 \left[\frac{\int_\Omega \varepsilon_c(\mathbf{r}) |\mathbf{e}_1(\mathbf{r})|^2 |\mathbf{e}_1(\mathbf{r})|^2 d\Omega}{\left(\int_\Omega \varepsilon_c(\mathbf{r}) |\mathbf{e}_1(\mathbf{r})|^2 d\Omega \right)^2}\right]. \label{eq:a12}
\end{equation}
To proceed, let
\begin{equation}
    \chi_{mn} = \frac{\int_\Omega \varepsilon_c(\mathbf{r}) |\mathbf{e}_m(\mathbf{r})|^2 |\mathbf{e}_n(\mathbf{r})|^2 d\Omega}{\left(\int_\Omega \varepsilon_c(\mathbf{r}) |\mathbf{e}_m(\mathbf{r})|^2 d\Omega \right)\left(\int_\Omega \varepsilon_c(\mathbf{r}) |\mathbf{e}_n(\mathbf{r})|^2 d\Omega \right)}
\end{equation}
denote the interaction coefficients \cite{Ge_2010}. Solving \eqref{eq:a12} for $\Gamma_1 e_c^{-2}|a_1|^2$ and into \eqref{eq:a11}, one can write an implicit equation for the interacting second modal threshold,
\begin{equation}
    d^{(2)}_{\textrm{int thr}} = d^{(2)}_{\textrm{thr}} \left(1 + \left(\frac{d^{(2)}_{\textrm{int thr}}}{d_{\textrm{thresh}}} -1\right) \frac{\chi_{21}}{\chi_{11}} \right),
\end{equation}
where $\chi_{11}$ denotes the self-saturation of the first mode (how much it depletes its own gain) and $\chi_{12}$ denotes the cross-saturation of mode 2 due to mode 1 (caused by the mode overlap). Finally, one can recover an explicit equation for the interacting thresholds of a second lasing mode as
\begin{equation}\label{eq:int}
    d^{(2)}_{\textrm{int thr}} = d^{(2)}_{\textrm{thr}} \left(\frac{1 - \chi_{21}/\chi_{11} } {1 - {d^{(2)}_{\textrm{thr}}\chi_{21}}/\left({d_{\textrm{thresh}}\chi_{11}}\right) }\right) .
\end{equation}
Note that this formula for the interacting threshold must be used with some caution, as without some care it can produce the misleading result of a negative interacting threshold. Instead, such a negative $d^{(2)}_{\textrm{int thr}}$ indicates that the mode never reaches threshold. To see, this, first note that by definition $d^{(2)}_{\textrm{thr}} > d_{\textrm{thresh}}$, and similarly, $\chi_{21} \le \chi_{11}$, as the interacting modal overlap between two modes cannot be greater than the interacting overlap of a mode with itself. If $\chi_{21} = 0$, i.e., the two modes do not spatially overlap, the interacting threshold of the second lasing mode is the same as its non-interacting value, $d^{(2)}_{\textrm{int thr}} = d^{(2)}_{\textrm{thr}}$. As $\chi_{21}$ is increased from zero, the ratio of $d^{(2)}_{\textrm{thr}} / d_{\textrm{thresh}}$ that yields a positive denominator becomes bounded. If, as a thought experiment, one could independently control $d^{(2)}_{\textrm{thr}}$ without altering $\chi_{21}$, increasing $d^{(2)}_{\textrm{thr}}$ above $d_{\textrm{thresh}}$ first yields a zero denominator, indicating a diverging $d^{(2)}_{\textrm{int thr}}$, before the interacting threshold becomes negative.Thus, within the SPA-SALT approximation,
\begin{equation}
    \frac{d^{(2)}_{\textrm{thr}}\chi_{21}}{d_{\textrm{thresh}}\chi_{11}} \ge 1 \implies  d^{(2)}_{\textrm{int thr}} \rightarrow \infty.
\end{equation}
This divergence arises from the SPA‑SALT approximation, which fixes the modal frequencies and spatial profiles. Under this assumption, a second mode cannot reshape its profile to access unused gain, so the interacting threshold formally diverges. Physically, this does not mean the threshold is infinite; it simply indicates that the second mode would require pumping far beyond the nonlinear regime where SPA‑SALT is valid. In reality, the threshold is extremely large but finite, supporting the conclusion that the system remains effectively single-mode under typical operating conditions. In \appref{sec:app_val} we use the interacting thresholds to see compare the lasing thresholds of the first and second lasing modes.

\subsection{Scalar residual for the SALT model}

Using the expression in \eqref{eq:a12} one can derive the lasing intensity for an arbitrary pumping strength above threshold $\Delta d = d-d_\text{thresh}$,
\begin{equation}\label{eq:a16}
    |a_1|^2 = \frac{\Delta d}{\Gamma_1 e_c^{-2} \chi_{11} d_\text{thresh}}\,,
\end{equation}
which is the amplitude expression that we consider in \appref{sec:app_B} close to threshold ($\Delta d \ll 1$) and in the weak pumping regime ($|a_1|^2 \ll 1$).

By using the expression for the lasing intensity derived in \eqref{eq:a16} together with \eqref{eq:D_E} in the single mode regime, it is possible to find the population inversion above threshold,
\begin{equation}
    D(\bvec{r})=   \left[1+  \frac{d-d_\text{thresh}}{\chi_{11} d_\text{thresh}} \left|\mathbf{e}_1(\bvec{r})\right|^2 \right]^{-1} D_0(\bvec{r}) d \, ,
\end{equation}
which can be directly inserted into  
\eqref{eq:SALT}. With this, we can define a normalized scalar residual measure based on the weak form of our finite element formulation that is a function of the pumping strength as
\begin{equation}\label{eq:res}
    r (d)= \frac{\int_\Omega|\nabla \times \bvec{E}_\text{r}(\bvec{r})|^2 -\left[\varepsilon_c(\bvec{r})-i D(\bvec{r},d)\right] \left(\omega_1 / c\right)^2\ |\bvec{E}_\text{r}(\bvec{r})|^2 \d\Omega}
 {\int_\Omega\ \left(\omega_1 / c\right)^2\varepsilon_c(\mathbf{r}) |\bvec{E}_\text{r}(\bvec{r})|^2 \d\Omega }\,,
\end{equation}
where we approximate the passive cavity mode with the reciprocal field in the high-$Q$ limit, while still considering the lasing frequency to be the real part of the cavity's resonance frequency ($\omega_L \approx \omega_1$). Thus, the residual will account for errors in differences between the frequency of the mode and the one used to solve the reciprocal problem, as well as the error introduced through the nonlinear hole burning term. In \appref{sec:app_val} we evaluate this residual for different values of the pumping strength $d$ to validate the high-$Q$ approximation at threshold and above, which allows to see how spatial hole burning affects the designs optimized close to threshold ($\Delta d = d - d_\text{thresh} \ll 1$).

\section{Lasing mode amplitude in the weak pumping regime}\label{sec:app_B}

Following a similar derivation to the single mode lasing amplitude in \appref{sec:app_SPA}, in this appendix we detail the calculation of the lasing amplitude in the high-$Q$, small amplitude, and small pumping strength limit. It is possible to expand \eqref{eq:gain_amp} for small amplitudes ($|a_1|^2 \ll 1$), including the zeroth and the first-order terms. This gives
\begin{equation}
 \Delta \varepsilon_\Im (\mathbf{r}) \approx  D_0(\mathbf{r}) d\left( 1 - |a_1|^2\,  \frac{ e_c^{-2}|\mathbf{e}_{1}(\mathbf{r})|^2}{\int_\Omega \varepsilon_c(\mathbf{r}) |\mathbf{e}_{1}(\mathbf{r})|^2 \d \Omega}\right)\,.
\end{equation}
Note that this first-order correction has the form of a Kerr nonlinearity, where the change in dielectric permittivity depends of on the electric-field intensity. In the high-$Q$ cavity limit, we can use \eqref{eq:pert}, where we use perturbation theory to calculate the shift in frequency of the mode, and substitute in the lasing condition in \eqref{eq:las_cond}. Using these expressions and assuming that we operate close to the lasing threshold where $d \approx d_\text{thresh}$ and $\Delta d = d - d_\text{thresh} \ll d_\text{thresh}$, the lasing mode intensity is given by
\begin{equation}
    |a_1|^2 \approx \frac{\Delta d}{ d_\text{thresh}}\frac{{\int_\Omega \varepsilon_c(\mathbf{r}) |\mathbf{e}_{1}(\mathbf{r})|^2 \d \Omega}\int_\Omega D_0(\mathbf{r})|\mathbf{e}_{1}(\mathbf{r})|^2 \,  \d \Omega} { e_c^{-2}\int_\Omega D_0(\mathbf{r}) |\mathbf{e}_{1}(\mathbf{r})|^4 \,  \d \Omega} \,,
\end{equation}
where $\Delta d $ denotes difference of the pumping strength $d$ with the threshold. (At large amplitudes, many additional nonlinear effects arise, including competition from additional lasing modes, eventually requiring a full nonlinear solver~\cite{EsterhazyLiu14}.)

\section{Lasing threshold and FOM with
steady-state diffusion}\label{sec:app_C}

In this appendix, we revisit the derivations of the laser properties (e.g., threshold, lasing mode intensity) and FOM, including steady-state diffusion effects through the C-SALT formalism \cite{csalt}. We introduce diffusion in the gain model in \eqref{eq:gain_diff}. By rewriting the lasing mode as a function of the passive cavity mode, we find
\begin{equation}\label{eq:gain_}
    \Delta \varepsilon_\Im(\mathbf{r}) =  \mathbb{S} \left( \mathbb{I} + |a_1|^2 \,\frac{e_c^{-2}|\mathbf{e}_{1}(\mathbf{r})|^2}{\int_\Omega \varepsilon(\mathbf{r}) |\mathbf{e}_{1}(\mathbf{r})|^2 \d \Omega} \mathbb{S} \right)^{-1} [D_0](\mathbf{r}) d\,.
\end{equation}

Similar to the case without diffusion, close to the lasing threshold, we assume a small amplitude \linebreak $(|a_1|^2\ll1)$, so we can expand the expression in a Taylor series around $|a_1|^2=0$. For zero amplitude the change in permittivity is $\Delta \varepsilon_\Im (\mathbf{r}) = \mathbb{S}[D_0] (\mathbf{r})d \,.$ Using this expression together with the lasing condition in \eqref{eq:las_cond}, gives an updated version of the threshold pumping strength that accounts for diffusion effects, resulting in \eqref{eq:pump_thresh_diff}.

By including the zero-th and the first-order terms in the expansion, we calculate the change in permittivity in the gain medium
\begin{equation}
 \Delta \varepsilon_\Im (\mathbf{r}) \approx \mathbb{S} \left( \mathbb{I} - |a_1|^2\,\frac{e_c^{-2}|\mathbf{e}_{1}(\mathbf{r})|^2}{\int_\Omega \varepsilon(\mathbf{r}) |\mathbf{e}_{1}(\mathbf{r})|^2 \d \Omega} \mathbb{S}\right)[D_0](\mathbf{r}) d\,.
\end{equation}
Following the same relations as in \appref{sec:app_B} gives an expression for the intensity of the lasing mode
\begin{equation}
    |a_1|^2 \approx \frac{\Delta d}{d_\text{thresh}} \frac{{\int_\Omega \varepsilon (\mathbf{r}) |\mathbf{e}_{1}(\mathbf{r})|^2 \d \Omega}\int_\Omega \mathbb{S} [D_0](\mathbf{r})|\mathbf{e}_{1}(\mathbf{r})|^2 \,  \d \Omega}{e_c^{-2}\int_\Omega \mathbb{S}\left[ |\mathbf{e}_{1}(\mathbf{r})|^2\, \mathbb{S} [D_0] \right] (\mathbf{r})|\mathbf{e}_{1}(\mathbf{r})|^2 \,  \d \Omega} \,.
\end{equation}
Using this expression together with the TCMT equations (\secref{sec:TCMT}), we obtain the FOM of \eqref{eq:eff_diff}.

\section{Validating the SPA-SALT approximations}\label{sec:app_val}
To validate the SPA-SALT approximation, we will use the derivations in \appref{sec:app_SPA}. Primarily, we will evaluate the two key approximations in our SPA-SALT-based optimization framework: that a single cavity mode will lase with a threshold significantly lower than other cavity modes, and that the electric field in the reciprocal problem accurately represents the single lasing mode at threshold and above. As an example, we will evaluate this by using the device optimized in 2d for a gain region with $\sigma=500$ nm (i.e., the cavity design shown in \figref{fig:nonlinear} (d) and \figref{fig:benchmark_fields}).

\subsection{Lasing threshold for a second lasing mode}

First, we identify the cavity modes closest to the optimized lasing mode. These modes are resonances with a maximum $Q$-factor of $Q\approx 200$ ($\approx$ 1-2 orders of magnitude lower than the lasing mode), which are spectrally separated by just a few nanometers. Given this spectral closeness, we approximate that the closest modes do not experience any line pulling effects ($\gamma_\perp \to \infty$, so that $\Gamma_1\to1$ and $\Gamma_2\to1$).
By using the expressions for the interacting threshold of a second lasing mode in \eqref{eq:int} and assuming $\omega_{L_2}=\omega_2$, we find that the value for the second interacting threshold for the closest modes diverge:  $d^{(2)}_{\textrm{int thr}}\to \infty$. This divergence arises from the SPA‑SALT approximation, which assumes that the modal frequencies and spatial profiles are fixed. Under this assumption, a second mode cannot adapt its profile to access unused gain, so the interacting threshold formally diverges. Physically, this does not imply an infinite threshold; rather, the second mode would require pumping far beyond the nonlinear regime where SPA‑SALT is valid, so the threshold is extremely large but finite. This means that within the SPA-SALT approximation a second mode will not reach threshold in the presence of modal interactions. This result thus supports the assumption that our system is dominated by a single lasing mode. Furthermore, the small modal overlap (quantified by $\chi_{21}/\chi_{11}$) ensures that even modes close in frequency are strongly suppressed by the first lasing mode, reinforcing single-mode operation. In a full SALT calculation allowing modal profile adaptation, the second mode could eventually turn on at very high pump, but this occurs well outside the parameter regime considered here.

\subsection{Linear response validation for high-$Q$ cavities}

Next, we validate the error between the linear response of the reciprocal system and the SALT model above the threshold by using the scalar residual measure defined in \eqref{eq:res}. The residual gives a scalar measure of the relative error between the linear solution and the SALT model, which can be used to validate the high-$Q$ approximation (i.e., that the reciprocal field is proportional to the cavity mode). 

\begin{figure*}[h!] 
    \centering
    \includegraphics[scale=1]{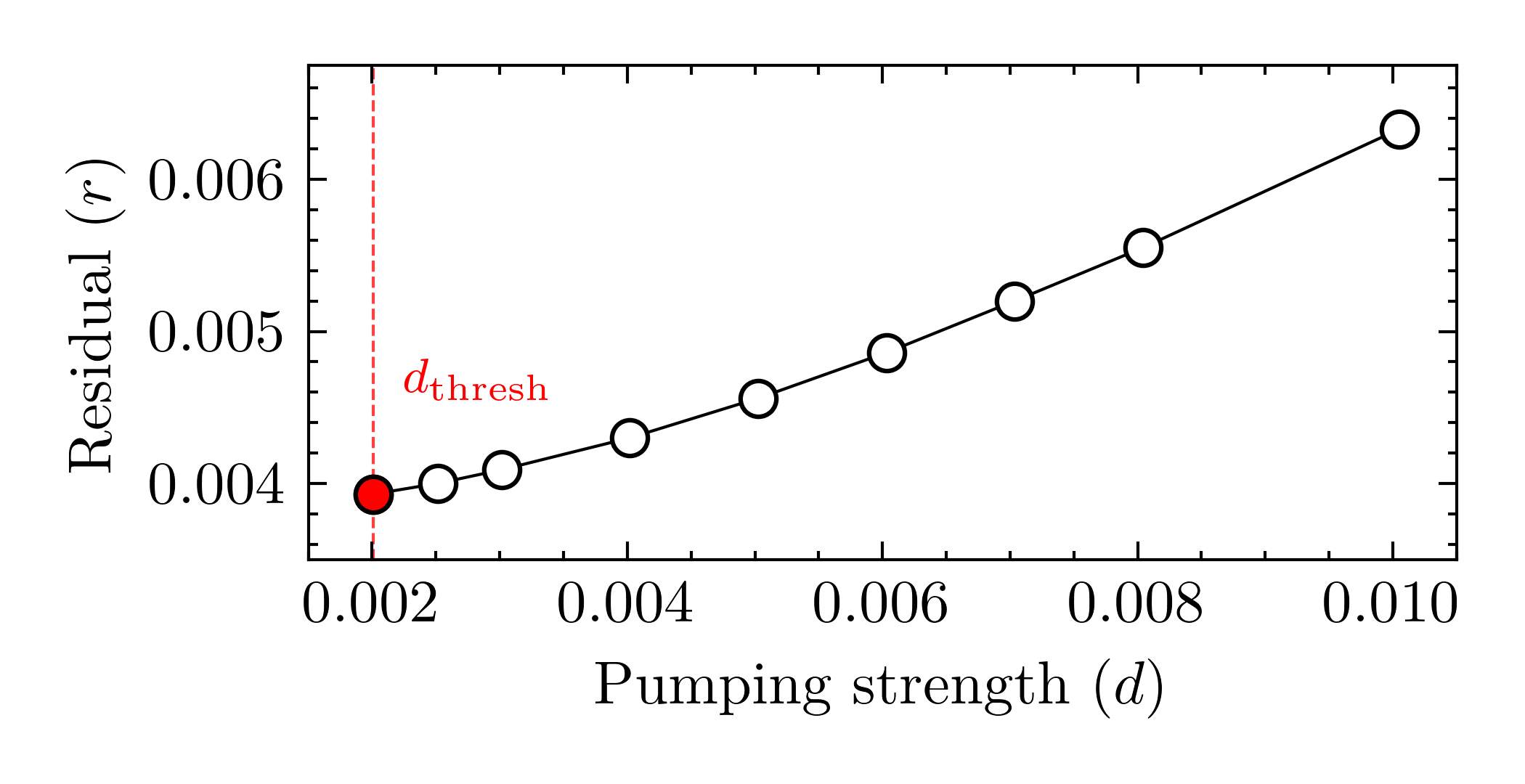} 
    \caption{Validation of the linear response high-$Q$ approximation at and above threshold. The residual as a function of pump strength $d$ for pumping strengths up to $d\approx 5 d_\text{thresh}$. The lasing threshold is marked in red at $d_\text{thresh}=2.01\cdot 10^{-3}$.}
    \label{res}
\end{figure*}

To show the validity of this approximation in \figref{res}, we plot the residual (\eqref{eq:res}) as a function of the scalar pumping strength ($d$). Close to the threshold ($d_\text{thresh}$), the residual is very low (a relative error of $\lessapprox 0.4 \%$), whereas as the pumping strength increases, the residual and the relative error start to increase. Nevertheless, even for a larger pumping strength of $d\approx 5 d_\text{thresh}$, the relative error is still smaller than $\lessapprox 1 \%$. We note that for even larger pumping strengths, the residual grows as a linear function, and we calculate that for a pumping strength of $d\approx 100 d_\text{thresh}$ the relative error is still $\approx 5 \%$. These results confirm the robustness and validity of the approximation for the inverse-designed high-$Q$ nanolaser cavities.


\medskip

\section{Simulation and optimization parameters}\label{sec:app_D}

In this section we detail the parameters used in the simulation and optimization of the nanolaser devices. The numerical model is discretized using the finite-element method~\cite{jin}, with first-order Nedelec elements~\cite{jin} for the frequency-domain electromagnetic problem and linear Lagrange elements for the steady-state exponentially damped diffusion problems.

The simulations are performed in COMSOL Multiphysics (version 6.2)~\cite{COMSOL} and the optimization is solved using the Globally Convergent Method of Moving Asymptotes (GCMMA)~\cite{GCMMA}. All computations are executed on the DTU Computing Center HPC cluster~\cite{DTU_DCC_resource}. 

\subsection*{Geometrical parameters}

\textbf{Two dimensional systems:}
In the 2d computational results (\secref{sec:benchmark}, \ref{sec:diffusion_results}) we consider a geometry based on a rectangular simulation domain with length $l=13.95\,\upmu\text{m}$ (along the $x$ direction) and width $w=4.55\,\upmu\text{m}$ (along the $y$ direction), where the rectangular design domain ($\Omega$) is centered at position $\bvec{r}_\text{0}=(1.1625\,\upmu\text{m}$, $0\,\upmu\text{m}$), and has side of length of $l_\Omega =1.5\lambda=2.325\,\upmu\text{m}$. The design domain is connected to a waveguide with width $w_\text{wg}= 500$~nm, which is excited at position $\bvec{r}_\text{src}=(-1\,\upmu\text{m}$, $0\,\upmu\text{m}$). 

\noindent\textbf{Three dimensional systems:}
In the 3d computational results (\secref{sec:3D}), we extend the previous geometry by extruding the two-dimensional design domain ($\Omega$) and waveguide out-of-plane (along the $z$-direction) to a height of $h_\Omega = 220$~nm, and placing the resulting structure on a silica substrate. The design domain is at the center of a simulation domain with total height of $h=1.5\lambda=2.325\,\upmu\text{m}$.

\subsection*{Numerical discretization}

In the 2d plane the design is meshed with a structured square grid with 40 nm side length, while the rest of the geometry is discretized using second-order elements on an unstructured triangular mesh with a side length of $40$ nm in the solid ($n_1$) and 155 nm in the air ($n_0$), so that the size corresponds to roughly $1/10$ in-material wavelengths. In the 3d simulations the waveguide and design domain are extruded with 5 elements out-of-plane, and the rest of the simulation domain is meshed with tetrahedral elements with side lengths corresponding to $1/10$ in-material wavelengths. Note that all reported performances and fields are evaluated using the final post processed design, discretized using an unstructured triangular (tetrahedral) body fitted mesh with 20 nm ($\sim \lambda/(20\,n_1)$) side length for the optimized structures.

\subsection*{Optimization parameters}

The main optimization parameter choices are the threshold value $\eta=0.5$ and filter radius $r_{\text{f}}=100$~nm. The geometric length scale constraints~\cite{lengthscale} are enforced using the parameters $c= 39$, and $\eta_e=0.75$ and $\eta_d=0.25$, as to ensure a lengthscale $\sim r_\mathrm{f}/2=50$ nm~\cite{meep}. The rest of the parameters, including the length scale error ($\epsilon$) needed to enforce the length scale constraints
~\cite{lengthscale}, are chosen as part of a continuation scheme, which is essential to enable the use of gradient-based methods and ensures a well-performing final binary
design. In the continuation scheme we vary the simulation parameters as a function of the iteration number to find a binary and well performing final design. We summarize those parameter choices in \tabref{tabcont}.

\begin{table}[h!]
\caption{Continuation scheme parameters in the topology optimization framework.}\label{tabcont}%
\begin{tabular}{@{}l|lllllllll@{}}
\toprule
	 \textbf{Continuation step} & $0$ & 1 & 2  & 3 & 4 & 5 & 6 & 7 & 8    \\
  \midrule
  Iteration & $0$ & 100 & 200  & 300 & 400 & 500 & 600 & 700 & 800    \\
  Threshold sharpness (\textbf{$\beta$}) & 2.5 & 5 & 7.5 & 10  & 15 & 25 & 50 & 75 & 100   \\
  Attenuation factor (\textbf{$\alpha^\prime$}) & 0.0 & 0.0 & 0.01  & 0.1 & 0.2 & 0.4 & 0.8 & 0.8 & 0.8   \\
   Artificial loss (\textbf{$\alpha$}) & $0.1$ & 0.05 & 0.025  & 0.01 &  $5 \cdot 10^{-3}$ & 0.0 & 0.0 & 0.0 & 0.0    \\
  Length scale error ($\epsilon$) & 1 & 1 & 1  & $10^{-3}$ & $5 \cdot 10^{-4}$ & $10^{-4}$ & $5 \cdot 10^{-5}$ & $ 10^{-5}$ & $5 \cdot 10^{-5}$
	\end{tabular}
\end{table}

\subsection*{Acknowledgements}
We thank A.D.~Stone for useful discussions. We gratefully acknowledge financial support from the Danish National Research Foundation through the NanoPhoton Center for Nanophotonics, grant number DNRF147. A.C.~acknowledges support from the Laboratory Directed Research and Development program at Sandia National Laboratories.
This work was performed in part at the Center for Integrated Nanotechnologies, an Office of Science User Facility operated for the U.S. Department of Energy (DOE) Office of Science.
Sandia National Laboratories is a multimission laboratory managed and operated by National Technology \& Engineering Solutions of Sandia, LLC, a wholly owned subsidiary of Honeywell International, Inc., for the U.S. DOE's National Nuclear Security Administration under Contract No. DE-NA-0003525.  S.G.J.~was supported in part by the Simons Foundation collaboration on Extreme Wave Phenomena, and by the U.S. Army Research Office (ARO) through the Institute for Soldier Nanotechnologies (ISN) under award no.~W911NF-23-2-0121.
The views expressed in the article do not necessarily represent the views of the U.S. DOE or the United States Government.
\medskip


%

\makeatletter
\renewcommand{\thesection}{\arabic{section}}%
\titleformat{\section}[block]
  {\normalfont\LARGE\bfseries} 
  {\thesection.}               
  {0.5em}{}
\makeatother

\bibliographystyle{ieeetr}
\bibliography{biblio}

\end{document}